\documentclass[5p]{elsarticle}
\usepackage[english]{babel}
\usepackage{graphicx}
\usepackage{amsmath, amssymb, upgreek}
\graphicspath{{Figs/}}
\DeclareGraphicsExtensions{.png}

\journal{Journal of Magnetism and Magnetic Materials}

\begin{document}

\begin{frontmatter}

\title{Power Loss for a Periodically Driven Ferromagnetic Nanoparticle in a Viscous Fluid: the Finite Anisotropy Aspects}

\author{T.~V.~Lyutyy\corref{cor_auth}}
\cortext[cor_auth]{Corresponding author}
\ead{lyutyy@oeph.sumdu.edu.ua}

\author{O.~M.~Hryshko}
\author{A.~A.~Kovner}

\address{Sumy State University, 2 Rimsky-Korsakov Street, UA-40007 Sumy, Ukraine}

\begin{abstract}
The coupled magnetic and mechanical motion of a ferromagnetic nanoparticle in a viscous fluid is considered within the dynamical
approach. The equation based on the total momentum conservation law is used for the description of the mechanical rotation,
while the modified Landau-Lifshitz-Gilbert equation is utilized for the description of the internal magnetic dynamics.
The exact expressions for the particles trajectories and the power loss are obtained in the linear approximation. The comparison with the results of other widespread approaches, such as the model of fixed particle and the model of rigid dipole, is performed. It is established that in the small oscillations mode the damping precession of the nanopartile magnetic moment is the main channel of energy dissipation, but the motion of the nanoparticle easy axis can significantly influence the value of the resulting power loss.
\end{abstract}


\end{frontmatter}


\section{Introduction}

The correct description of a ferromagnetic nanoparticle dynamics in a viscous carrier fluid is a key to understanding the ferrofluid dynamics for all possible applications. Up to now, for ferrofluids composed of small enough nanoparticles, the response to a time-periodic magnetic field was considered firstly within the concept of complex magnetic susceptibility, which is well described in \cite{Rosensweig2002370}. However, when the nanoparticle magnetic energy is comparable with the thermal one, the response of a nanoparticle will be based mainly on the individual trajectories of each nanoparticle. For example, the regular viscous rotation is considered as the main energy dissipation channel for large enough nanoparticles driven by an external alternating field \cite{andr2006magnetism}. This gives reason to believe that the analytical description of the single nanoparticle motion is demanded.

Two components of the nanoparticle dynamics should be considered simultaneously for the trajectory description: 1) the mechanical rotation (or the so-called spherical motion) of a nanoparticle with respect to a viscous fluid, 2) the internal motion of the nanoparticle magnetization in the framework of the crystal lattice. Since the simultaneous description is faced with some difficulties, two approximations are utilized instead: 1) the rigid dipole approach \cite{0038-5670-17-2-R02}, when the nanoparticle magnetic moment is supposed to be locked in the nanoparticle crystal lattice, 2) the fixed nanoparticle approach \cite{PhysRev.130.1677}, when a nanoparticle is assumed to be immobilized because of the rigid bound with a media carrier. Despite the restrictions, both approaches are widely used for the description of the response to an alternating field of a ferromagnetic particle in a viscous fluid, including the power loss calculation problem, which is closely related to the magnetic fluid hyperthermia for cancer therapy \cite{Jordan1999413, 0022-3727-36-13-201}. Thus, the model of rigid dipole was applied successfully for the dynamical and stochastic approximations: the power loss was found for a circularly-polarized \cite{Raikher2011, PhysRevE.83.021401, 7753812} and a linearly-polarized \cite{0953-8984-15-23-313, 7753812} fields. The effective Langevin equation and the key characteristics of the rotational dynamics were established in \cite{PhysRevE.92.042312}. The power loss calculation within the fixed nanoparticle model, where only a damping precession of the magnetic moment is taken into account, was performed in \cite{PhysRevE.86.061404, PhysRevE.93.012607, PhysRevB.91.054425}. Finally, this problem was investigated in \cite{PhysRevB.85.045435, PhysRevB.89.014403} for the nanoparticles ensemble.

The coupled dynamics of a nanoparticle cannot be described by a simple superposition of these two types of motion because of the essential changes in the equations of motion. The coupled motion of the particle magnetic moment and the whole particle was firstly described in \cite{Cebers1975}. Despite this, the discussion about the basic equations of motion in the case of the coupled dynamics is continued till now \cite{Mamiya2011,doi:10.1063/1.4737126,doi:10.1063/1.4937919,Usov2015339}. It is especially important in the context of a ferrofluid heating by an alternating field, when both these types of motion can produce heat. One of the first successful attempts concerning the energy absorption description was reported in \cite{0022-3727-39-22-002}. There the power loss was obtained in the dynamical approximation by linearization of the Lagrangian equation in some specific cases. But within this approach the equations of motion were not used. The study of the forced coupled dynamics in a circularly-polarized magnetic field using the simplified equations of motion was presented in \cite{PhysRevE.87.062318}, but the energy absorption problem was not considered. The power loss was calculated in recent studies \cite{Mamiya2011, doi:10.1063/1.4737126}. Unfortunately, the correct explicit form of the equations of motion was not applied there that facilitates the discussion about the basic model equations \cite{doi:10.1063/1.4937919,Usov2015339}. And only recently the essential progress in the description of energy absorption by a viscously coupled nanoparticle with a finite anisotropy was achieved \cite{PhysRevB.95.134447}. Here the microwave absorption spectra was investigated using the linear response approach. But the viscous term as well as in \cite{Cebers1975} was not taken into account that motivates further research.

Therefore, we use the correct equations set, presented in \cite{doi:10.1063/1.4937919} to investigate the nanoparticle response to an external alternating field. Absorption of the field energy, which further is transformed into heating, is in our main focus. In particular, the influence of the easy axis mobility on the resonance dependencies of the power loss on the field frequency is examined. Then, we consider the results obtained for the same conditions within the fixed nanoparticle and the rigid dipole approximations. In this way we reveal the role of both the viscous rotation of a whole particle and the damping precession of its magnetic moment inside in the energy dissipation process. We make a conclusion about the complex character of the coupled dynamics and indistinguishability of the contribution of each type of motion into the mutual heating in the dynamical approximation.

\section{Description of the Model}

Let us consider a uniform spherical single-domain ferromagnetic nanoparticle of radius $R$, magnetization ($\mathbf{M}$, $\mathbf{|M|}= M = \mathrm{const}$) and density $\rho$. This particle performs the spherical motion (or motion with the fixed center of mass) with respect to a fluid of viscosity $\eta$. Then, we assume that the nanoparticle is driven by the external time-periodic field of the following type:

\begin{equation}
    \mathbf{H}(t) = \mathbf{e}_{x} H\cos(\Omega t) + \mathbf{e}_{y}\sigma H\sin(\Omega t),
    \label{eq:def_h}
\end{equation}

where $\mathbf{e}_x$, $\mathbf{e}_y$ are the unit vectors of the Cartesian coordinates, $H$ is the field amplitude, $\Omega$ is the field frequency, $t$ is the time, and $\sigma$ is the factor which determines the polarization type ($\sigma=\pm1$ corresponds to the circularly polarized field, $0<|\sigma|<1$ corresponds to the elliptically polarized field, and $\sigma=0$ corresponds to the linearly polarized field). Both the interaction with the viscous media and the damping precession of $\mathbf{M}$ inside the particle lead to dissipation of the nanoparticle energy and further heating of the surrounding environment. These losses are compensated by the energy absorption of the external field of type (\ref{eq:def_h}) and can be characterized by the dimensionless power loss per period calculated as \cite{PhysRevB.91.054425}

\begin{equation}
    q =  \frac{\Omega} {2\pi M H_a} \int_{0}^{\frac{2\pi} {\Omega}} dt\, \mathbf{H}\frac{\partial\mathbf{M}} {\partial t}.
    \label{eq:def_q}
\end{equation}

\subsection{Equations of Motion}
Three approaches will be considered: 1) the model of viscously coupled nanoparticle with a finite anisotropy (FA-model),  2) the model of fixed particle (FP-model), 3) the model of rigid dipole (RD-model). Let us start with the first one as more important and novel. As follows from \cite{doi:10.1063/1.4937919}, the coupled magnetic dynamics and mechanical motion in the deterministic case obey a pair of coupled equations

\begin{equation}
\begin{array}{lcl}
    \dot{\mathbf{n}}= \boldsymbol{\upomega}\times \mathbf{n}, \\
    [2pt]
    J \dot{\boldsymbol{\upomega}}=\gamma^{-1}V\dot{\mathbf{M}} + V\mathbf{M}\times \mathbf{H}- 6\eta V\boldsymbol{\upomega},\\
    \label{eq:Main_Eq_FA_n}
\end{array}
\end{equation}

where $\mathbf{n}$ is the unit vector which determines the anisotropy axis direction, $\boldsymbol{\upomega}$ is the angular velocity of the particle, $J(= 8\pi \rho R^5/15)$ is the moment of inertia of the particle, $\gamma$ is the gyromagnetic ratio, $V$ is the nanoparticle volume, and dots over symbols represent derivatives with respect to time.

In fact, the first equation in the set (\ref{eq:Main_Eq_FA_n}) is the condition of spherical motion for a rigid body, and the second one is the classical torque equation, where the first term constitutes the main difference from the same equation for the FP-model.
This term is originated from the motion of magnetization inside the nanoparticle with respect to its crystal lattice. In turn, the magnetization dynamics is described by the modified Landau-Lifshitz-Gilbert (LLG) equation

\begin{equation}
    \mathbf{\dot{M}}=-\gamma  \mathbf{M}\times \mathbf{H}_{eff} +\alpha {M}^{-1}\left( \mathbf{M}\times  \mathbf{\dot{M}}-\boldsymbol{\upomega}\times \mathbf{M} \right),
    \label{eq:Main_Eq_FA_m}
\end{equation}

where $\alpha$  is the damping parameter, $\mathbf{H}_{eff}$ is the effective magnetic field, which accounts the uniaxial anisotropy field of magnitude $H_a$

\begin{equation}
    \mathbf{H}_{eff}=\mathbf{H}+{{H}_{a}}{{M}^{-1}}\left( \mathbf{Mn} \right)\mathbf{n}.
    \label{eq:H_eff}
\end{equation}

Here, the difference from the original LLG equation consists in the term which is proportional to $\mathbf{M} \times \boldsymbol{\upomega}\times \mathbf{M} $ that excludes the component of $\mathbf{M}$ rotating together with the crystal lattice. As a rule, the inertia term in (\ref{eq:Main_Eq_FA_n}) can be neglected even for large enough nanoparticles ($R > 20~\mathrm{nm}$) in a wide frequency range. Therefore, for further analysis we transform the equations of motion (\ref{eq:Main_Eq_FA_n}) and (\ref{eq:Main_Eq_FA_m}) into the convenient form

\begin{equation}
\begin{array}{lcl}
    \mathbf{\dot{n}}=M {H}_{a}\left[\mathbf{\dot{m}}\times \mathbf{n}/\Omega_{r}+\left(\mathbf{m}\times \mathbf{h} \right)\times \mathbf{n}\right]/6\eta,\\
    [2pt]
    \mathbf{\dot{m}}(1+\beta)=-\Omega_{r}\mathbf{m}\times \mathbf{h}_{eff}^{1}+\alpha \mathbf{m}\times \mathbf{\dot{m}},
    \label{eq:Red_Eq_FA}
\end{array}
\end{equation}

where $\Omega_r = \gamma H_a$ is the ferromagnetic resonance frequency, $\beta =\alpha M/6\gamma \eta$,

\begin{equation}
    \mathbf{h}_{eff}^{1}=\left(\mathbf{e}_{x} h\cos{\Omega t} +\mathbf{e}_{y}h\sigma \sin{\Omega t} \right)\left(1+\beta  \right)+\left(\mathbf{mn} \right)\mathbf{n},
    \label{eq:h_eff_1}
\end{equation}

and, finally, $\mathbf{m} = \mathbf{M}/M$, $h = H/H_a$ are the dimensionless magnetic moment and filed amplitude, respectively.

The FP-model is described by the well-known LLG equation

\begin{equation}
    \mathbf{\dot{M}}=-\gamma \mathbf{M}\times \mathbf{H}_{eff} +\alpha {M}^{-1}\mathbf{M}\times  \mathbf{\dot{M}}
    \label{eq:Main_Eq_FP_m}
\end{equation}

or in the dimensionless form

\begin{equation}
    \mathbf{\dot{m}}=-\Omega_r \mathbf{m}\times \mathbf{h}_{eff} +\alpha \mathbf{m}\times \mathbf{\dot{m}},
    \label{eq:Red_Eq_FP_m}
\end{equation}

where $\mathbf{h}_{eff}=\mathbf{H}_{eff}/H_a$.

Finally, the RD-model is described by the set of equations similar to the Eqs.~(\ref{eq:Main_Eq_FA_n}), but without the term proportional to $\dot{\mathbf{M}}$

\begin{equation}
\begin{array}{lcl}
    \dot{\mathbf{n}}= \boldsymbol{\upomega}\times \mathbf{n}, \\
    [2pt]
     J \dot{\boldsymbol{\upomega}} = VM \mathbf{n}\times \mathbf{H}- 6\eta V\boldsymbol{\upomega}.\\
    \label{eq:Main_Eq_FM_n}
\end{array}
\end{equation}

When the inertia momentum is neglected, Eqs.~(\ref{eq:Main_Eq_FM_n}) are transformed into a simple form

\begin{equation}
    \dot{\mathbf{n}}= - \Omega_{cr}\mathbf{n}\times\left( \mathbf{n}\times \mathbf{h}\right),
    \label{eq:Red_Eq_FM_n}
\end{equation}

where $\Omega_{cr}=M H_{a}/6 \eta$ is the characteristic frequency of the uniform mechanical rotation.

\subsection{Validation of the Dynamical Approximation}

The used systems of equations are valid if thermal fluctuations do not significantly influence the obtained trajectories. There are two principal issues in this regard needed to be considered. Firstly, the magnetic energy should be much larger than the thermal energy, or $ \Gamma \gg 1$, where $\Gamma = MHV/(k_{\mathrm{B}} T)$, $T$ is the thermodynamic temperature, $k_{\mathrm{B}}$ is the Boltzmann constant. In this case, primarily small deviations  from the dynamical trajectories take place. Secondly, the requirement to the relaxation time $\tau_{\mathrm{N}}$ exists. Here, relaxation time is the time, during which the rare, but large fluctuations can occur. When the period of an external field is much smaller than the relaxation time, or $\Omega^{-1}\ll\tau_{\mathrm{N}}$, the probability of such fluctuation is negligible, and the dynamical approach remains valid. Following Brown \cite{PhysRev.130.1677}, the relaxation time $\tau_{\mathrm{N}}$ can be found as $\tau_{\mathrm{N}}= (\Gamma/\pi)^{-1/2}\exp(\Gamma)(2\alpha \gamma H)^{-1}$. Both these factors together impose the requirements to the nanoparticle size and values of the field frequency and amplitude. For example, $\Gamma \approx 11.9$ for the real nanoparticles of maghemite \cite{C3RA45457F} with the following parameters: average radius $R=20~\mathrm{nm}$, $H_a = 910~\mathrm{Oe}$, $M = 338~\mathrm{G}$, temperature $T=315~\mathrm{K}$ and external field amplitude $H = 0.05 H_{a}$. Then, the frequency should to be larger than $\tau_{\mathrm{N}}^{-1}$, which for the parameters stated above and $\alpha=0.05$ is equal to $\Omega_{\mathrm{N}}\approx 1.11 \cdot 10^3~\mathrm{Hz}$.

These conditions are sufficient for the FP-model. But when we consider the mechanical rotation in addition to the magnetic dynamics within the FA-model, one needs to take into account the conditions of stable spherical motion. The significant changes in the angular coordinates can occur due to thermal excitation, when the observation time is much more than the Brownian relaxation time $\tau_{\mathrm{B}} = 3 \eta V/(k_{\mathrm{B}} T)$ \cite{Raikher_1994}. It imposes the existence of another characteristic frequency  $\Omega_{\mathrm{B}} = \tau_{\mathrm{B}}^{-1} = k_{\mathrm{B}} T/(3 \eta V)$. For the above-mentioned maghemite nanoparticles of radius $R=20~\mathrm{nm}$ and water at temperature of $T=315~\mathrm{K}$ and viscosity of $\eta=0.006~\mathrm{P}$ this frequency is equal to $\Omega_{\mathrm{B}} \approx 2.26 \cdot 10^5~\mathrm{Hz}$. One more requirement to the frequency arises from the condition which represents the validity of the Stokes approximation for the friction momentum \cite{frenkel1955kinetic}: $\mathrm{Re} = \rho_{l} \Omega_{\mathrm{S}} R^{2}/ \eta \sim 10$. Here $\mathrm{Re}$ is the so-called Reynolds number, $\rho_{l}$ is the liquid density, $\Omega_{\mathrm{S}}$ is the corresponding characteristic frequency, which defines the upper limit of the field frequency. Straightforward calculations give $\Omega_{\mathrm{S}} \sim 10^{12}~\mathrm{Hz}$ in our case. Summarizing, one can obtain that $\mathrm{max}[\Omega_{\mathrm{B}}, \Omega_{\mathrm{N}}] \ll \Omega \ll \Omega_{\mathrm{S}}$. Therefore, the frequency interval, where the dynamical approach is valid for the calculation, is $ \Omega = (10^{5}-10^{12})~\mathrm{Hz}$ which includes the frequencies acceptable for the magnetic fluid hyperthermia method.

Finally, the conditions of using the RD-model include all the stated above for the FA-model and contain additionally the requirement to the field amplitudes, which should be much smaller than the effective anisotropy field ($H \ll H_a$). The last inequality satisfies the above calculations and corresponds to the limitations of the linear approximation utilized for the processing of the equations of motion.

The importance of the dynamical approximation is not restricted by its validity in a certain interval of the system parameters. The dynamical approximation reveals the main microscopic mechanisms of the ferrofluid sensitivity to external fields. In this way we can estimate the upper limits of such important performance criteria as the magnetic susceptibility or the power loss. It is very important in a light of fictionalization of ferrofluids and creation of the properties demanded in the applications.

\section{Results}

\begin{figure}
    \centering
    \includegraphics[width=0.6\linewidth] {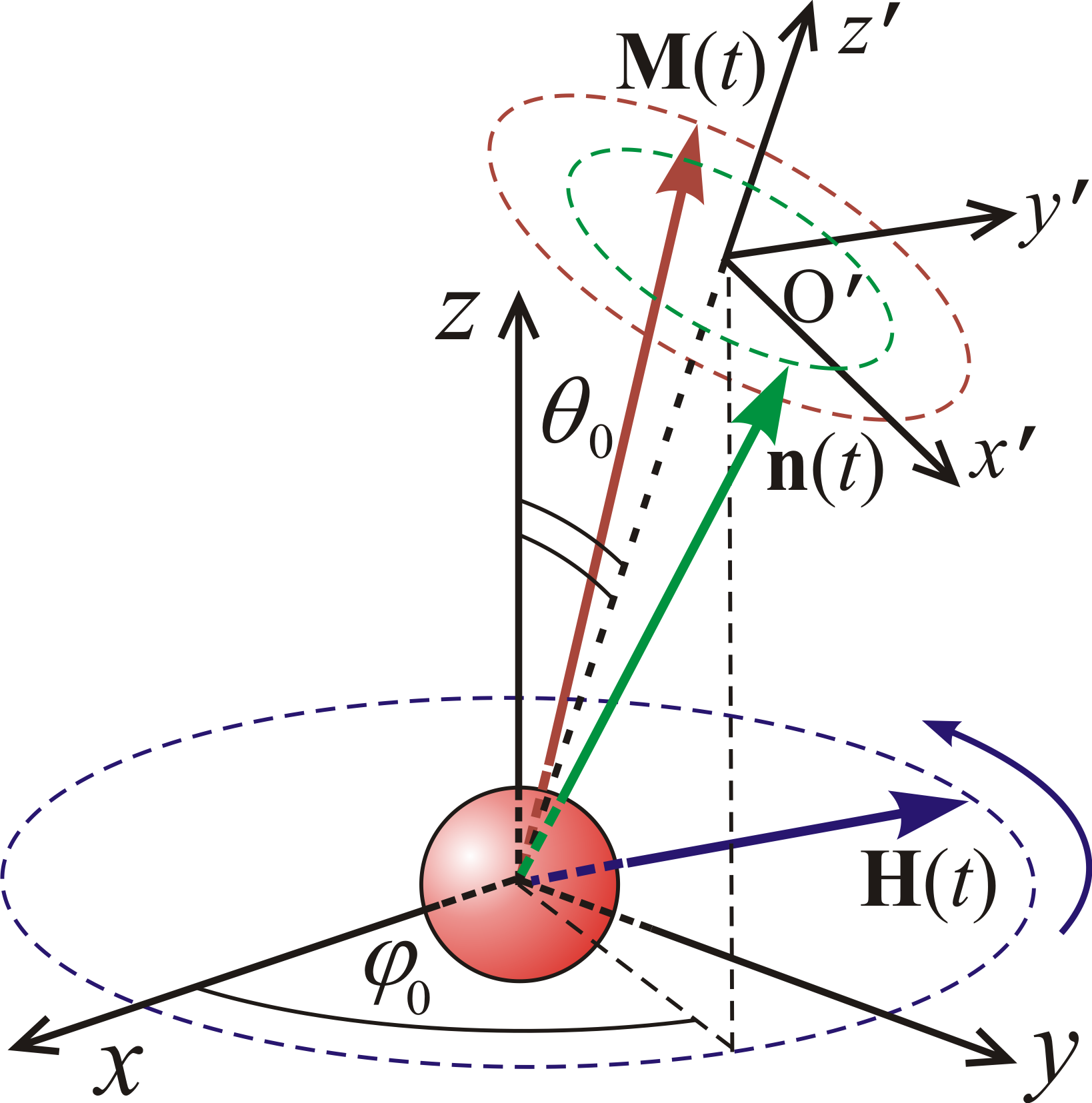}
    \caption {\label{fig:model} (Color online)
    Schematic representation of the model and the coordinate systems.}
\end{figure}

The solution of the set of equations (\ref{eq:Red_Eq_FA}), (\ref{eq:Red_Eq_FM_n}) and (\ref{eq:Red_Eq_FP_m}) can be found in the linear approximation for the small oscillations mode. In this mode, vectors $\mathbf{m}$ and $\mathbf{n}$ are rotated in a small vicinity around the initial position of the easy axis which, in turn, is defined by the angles $\theta_0$ and $ \varphi_0$ (see Fig.~\ref{fig:model}). This takes place for small enough field amplitudes ($h \ll 1$). The linearization procedure used here is similar to the reported in \cite{PhysRevB.91.054425} and consists in the following. Let us introduce a new coordinate system $x' y' z'$ in the way depicted in Fig.~\ref{fig:model}. Actually, it is rotated with respect to the laboratory system $xyz$ by the angles $\theta_0$ and $ \varphi_0$. In this new coordinate system, vectors $\mathbf{m}$ and $\mathbf{n}$ can be represented in the linear approximation as

\begin{equation}
\mathbf{m}=\mathbf{e}_{x'}m_{x'}+\mathbf{e}_{y'}m_{y'}+\mathbf{e}_{z'},
\label{eq:m_lin_gen_sol}
\end{equation}

\begin{equation}
\mathbf{n}=\mathbf{e}_{x'}n_{x'}+\mathbf{e}_{y'}n_{y'}+\mathbf{e}_{z'},
\label{eq:m_lin_gen_sol}
\end{equation}

where $\mathbf{e}_{x'}, \mathbf{e}_{y'}, \mathbf{e}_{z'}$ are the unit vectors of the coordinate system $x' y' z'$.
In this system, the external field (\ref{eq:def_h}) can be written using the rotation matrix as

\begin{equation}
\begin{array}{lcl}
    {\textbf{h}}' = \textbf{C}\cdot
        \left(
        \begin{array}{c}
            h \cos \Omega t \\
            \sigma h \sin \Omega t\\
            0 \\
        \end{array}
    \right),
    \label{eq:h_C}
\end{array}
\end{equation}

\begin{equation}
    {\textbf{C}} =
        \left(
        \begin{array}{lcr}
            \cos\theta_0 \cos\varphi_0 & \cos\theta_0 \sin\varphi_0 & - \sin\theta_0 \\
            -\sin\varphi_0  & \cos\varphi_0 & 0   \\
            \sin\theta_0 \cos\varphi_0 &  \sin\theta_0 \sin\varphi_0 & \cos\theta_0 \\
        \end{array}
        \right),
        \label{eq:C}
\end{equation}

\begin{equation}
\begin{array}{lcl}
    {\textbf{h}}' =
        \left(
        \begin{array}{l}
            h \cos\theta_0 \cos\varphi_0 \cos \Omega t + \sigma h \cos\theta_0 \sin\varphi_0 \sin \Omega t\\
            - h \sin\varphi_0 \cos \Omega t + \sigma h \cos\varphi_0 \sin \Omega t\\
            h \sin\theta_0 \cos\varphi_0 \cos \Omega t + \sigma h \sin\theta_0 \sin\varphi_0 \sin \Omega t \\
        \end{array}
    \right).
    \label{eq:h_prime}
\end{array}
\end{equation}

All the above allows to analyze the features of the nanoparticle response to the external field (\ref{eq:def_h}) for three approximations in an uniform manner. The analytical solutions obtained below describe the principal difference between the coupled and separated motion of the magnetic moment and the whole particle that constitutes our main results.

\subsection{Coupled Oscillations of the Easy Axis and the Magnetic Moment}

We start from the most complicated, however, the most interesting case: the case when both the mechanical rotation and internal magnetic dynamics occur simultaneously, or the FA-model. Using (\ref{eq:h_prime}), assuming $n_{x'}, n_{y'}, m_{x'}, m_{y'} \sim h$, and neglecting all the nonlinear terms with respect to $h$, we, finally, derive from (\ref{eq:Red_Eq_FA}) the linearized system of equations for $\textbf{m}$ and $\textbf{n}$ in the following form:

\begin{equation}
\begin{array}{l}
    \dot{n}_{x'} = M H_{a}\left( \dot{m}_{y'}/\Omega_{r} + h_{x'}\right)/{6\eta},\\
    [2pt]
    \dot{n}_{y'} = - M H_a \left( \dot{m}_{x'}/\Omega_{r}  -h_{y'}\right)/{6\eta},\\
    [2pt]
    (1 + \beta)\dot{m}_{x'} = - \Omega_{r} \left(m_{y'} - h_{y'} - n_{y'}\right) - \alpha \dot{m}_{y'},\\
    [2pt]
    (1 + \beta)\dot{m}_{y'} =  \Omega_{r} \left(m_{x'} - h_{x'} - n_{x'}\right) - \alpha \dot{m}_{x'}.\\
\end{array}
 \label{eq:Red_Eq_FA_lin}
\end{equation}

Solution of this set of linear equations can be written in the standard form

\begin{equation}
\begin{array}{lcl}
    n_{x'}= a_{n}\cos \Omega t + b_{n}\sin \Omega t, \\
    [2pt]
    n_{y'}= c_{n}\cos \Omega t + d_{n}\sin \Omega t, \\
    [2pt]
    m_{x'}= a_{m}\cos \Omega t + b_{m}\sin \Omega t, \\
    [2pt]
    m_{y'}= c_{m}\cos \Omega t + d_{m}\sin \Omega t, \\
    [2pt]
\end{array}
\label{eq:FA_lin_gen_sol}
\end{equation}

where $a_n$, $b_n$, $c_n$, $d_n$, $a_m$, $b_m$, $c_m$, and $d_m$ are the constant coefficients which should be defined. Substituting (\ref{eq:FA_lin_gen_sol}) into (\ref{eq:Red_Eq_FA_lin}) and using the linear independence of the trigonometric functions, we obtain the system of linear algebraic equations for the coefficients corresponding to $\mathbf{m}$

\begin{equation}
\begin{array}{rcl}
    (1 + \beta) \tilde{\Omega}a_{m} \!\!&=&\!\! d_{m}  + \delta b_{m} - \alpha \tilde{\Omega} c_{m} - A_{m}, \\
    [2pt]
    (1 + \beta) \tilde{\Omega}b_{m} \!\!&=&\!\! -c_{m} - \delta a_{m} - \alpha \tilde{\Omega} d_{m} + B_{m}, \\
    [2pt]
    (1 + \beta) \tilde{\Omega}c_{m} \!\!&=&\!\! - b_{m} + \delta d_{m} + \alpha \tilde{\Omega} a_{m} - C_{m}, \\
    [2pt]
    (1 + \beta) \tilde{\Omega}d_{m} \!\!&=&\!\!  a_{m} - \delta c_{m} + \alpha \tilde{\Omega} b_{m} + D_{m} \\
\end{array}
\label{eq:FA_alg_gen_eq}
\end{equation}

and the explicit expressions for the coefficients corresponding to $\mathbf{n}$

\begin{equation}
    \begin{array}{lcl}
    a_{n} = \delta c_{m} - \sigma \tilde{\Omega}^{-1} h \cos\theta_{0}\sin\varphi_{0},\\
    [2pt]
    b_{n} = \delta d_{m} + \tilde{\Omega}^{-1} h \cos\theta_{0}\cos\varphi_{0},\\
    [2pt]
    c_{n} = -\delta a_{m} - \sigma \tilde{\Omega}^{-1} h \cos\varphi_{0},\\
    [2pt]
    d_{n} = - \delta b_{m} - \tilde{\Omega}^{-1} h \sin\varphi_{0}.\\
\end{array}
\label{eq:FA_lin_gen_coef_n}
\end{equation}

Here $\tilde{\Omega} = \Omega/\Omega_{r}$, $\delta = \beta /\alpha$ and

\begin{equation}
    \begin{array}{lcl}
    A_{m}= \sigma h (1 +\beta) \cos\varphi_{0}  - \tilde{\Omega}^{-1} h \sin\varphi_{0},\\
    [2pt]
    B_{m}= - h (1 +\beta) \sin\varphi_{0}  - \sigma \tilde{\Omega}^{-1} h \cos\varphi_{0},\\
    [2pt]
    C_{m}= - \sigma h (1 +\beta) \cos\theta_{0}\sin\varphi_{0}  - \tilde{\Omega}^{-1} h \cos\theta_{0}\cos\varphi_{0},\\
    [2pt]
    D_{m}= - h (1 +\beta) \cos\theta_{0}\cos\varphi_{0}  + \sigma \tilde{\Omega}^{-1} h \cos\theta_{0}\sin\varphi_{0}.\\
\end{array}
\nonumber
\end{equation}

From (\ref{eq:FA_alg_gen_eq}) one straightforwardly obtains the unknown constants $a_m$, $b_m$, $c_m$, and $d_m$ as follows

\begin{equation}
\begin{array}{lcl}
    a_m \!\!&=&\!\! Z^{-1} \left[\tilde{\Omega}_1 D_m + \tilde{\Omega}_2 B_m + \tilde{\Omega}_3 C_m + \tilde{\Omega}_4 A_m \right],\\
    [2pt]
    b_m \!\!&=&\!\! Z^{-1} \left[\tilde{\Omega}_1 C_m + \tilde{\Omega}_2 A_m -\tilde{\Omega}_3 D_m -\tilde{\Omega}_4 B_m \right], \\
    [2pt]
    c_m \!\!&=&\!\! Z^{-1} \left[-\tilde{\Omega}_1 B_m + \tilde{\Omega}_2 D_m -\tilde{\Omega}_3 A_m + \tilde{\Omega}_4 C_m \right],\\
    [2pt]
    d_m \!\!&=&\!\! Z^{-1} \left[-\tilde{\Omega}_1 A_m + \tilde{\Omega}_2 C_m + \tilde{\Omega}_3 B_m -\tilde{\Omega}_4 D_m \right], \\
[2pt]
\end{array}
\label{eq:FA_lin_gen_coef_m}
\end{equation}

where

\begin{equation}
    \begin{array}{lcl}
   Z \!\!&=&\!\! \tilde{\Omega}^4 \alpha^4+2 \tilde{\Omega}^4 \alpha^2 \beta^2+\tilde{\Omega}^4 \beta^4 + 4 \tilde{\Omega}^4 \alpha^2 \beta+\\
    \!\!&+&\!\! 4 \tilde{\Omega}^4 \beta^3 + 2 \tilde{\Omega}^4 \alpha^2 + 6 \tilde{\Omega}^4 \beta^2 - 2 \tilde{\Omega}^2 \alpha^2 \delta^2+\\
    \!\!&+&\!\! 2 \tilde{\Omega}^2 \beta^2 \delta^2 + 4 \tilde{\Omega}^4 \beta + 8 \tilde{\Omega}^2 \alpha \beta \delta +\\
    \!\!&+&\!\! 4 \tilde{\Omega}^2 \beta \delta^2 + \tilde{\Omega}^4 + 2 \tilde{\Omega}^2 \alpha^2 + 8 \tilde{\Omega}^2 \alpha \delta-\\
    \!\!&-&\!\! 2 \tilde{\Omega}^2 \beta^2 + 2 \tilde{\Omega}^2 \delta^2+\delta^4 - 4 \tilde{\Omega}^2 \beta - 2 \tilde{\Omega}^2+\\
    \!\!&+&\!\! 2 \delta^2 + 1,\\
    \label{eq:FA_lin_gen_Z}
    \end{array}
\end{equation}

\begin{equation}
    \begin{array}{lcl}
   \tilde{\Omega}_1 \!\!&=&\!\! -\tilde{\Omega}^2 \alpha^2 - 2 \tilde{\Omega}^2 \alpha \beta \delta -2 \tilde{\Omega}^2 \alpha \delta +\\
    \!\!&+&\!\! \tilde{\Omega}^2 \beta^2 + 2 \tilde{\Omega}^2 \beta + \tilde{\Omega}^2 - \delta^2 - 1,\\
    [2pt]
    \tilde{\Omega}_2 \!\!&=&\!\! - \tilde{\Omega}^2 \alpha^2 \delta + 2 \tilde{\Omega}^2 \alpha \beta + 2 \tilde{\Omega}^2 \alpha + \\
    \!\!&+&\!\! \tilde{\Omega}^2 \beta^2 \delta + 2 \tilde{\Omega}^2 \beta \delta + \tilde{\Omega}^2 \delta + \delta^3 + \delta,\\
    [2pt]
    \tilde{\Omega}_3 \!\!&=&\!\! \tilde{\Omega}^3 \alpha^3 + \tilde{\Omega}^3 \alpha \beta^2 + 2 \tilde{\Omega}^3 \alpha \beta +\\
    \!\!&+&\!\! \tilde{\Omega}^3 \alpha - \tilde{\Omega} \alpha \delta^2 + \tilde{\Omega} \alpha + 2 \tilde{\Omega} \beta \delta +\\
    \!\!&+&\!\! 2 \tilde{\Omega} \delta,\\
    [2pt]
    \tilde{\Omega}_4 \!\!&=&\!\! - \tilde{\Omega}^3 \alpha^2 \beta - \tilde{\Omega}^3 \alpha^2 - \tilde{\Omega}^3 \beta^3 -\\ \!\!&-&\!\! 3 \tilde{\Omega}^3 \beta^2 - 3 \tilde{\Omega}^3 \beta - \tilde{\Omega}^3 - 2 \tilde{\Omega} \alpha \delta -\\ \!\!&-&\!\! \tilde{\Omega} \beta \delta^2 + \tilde{\Omega} \beta - \tilde{\Omega} \delta^2 + \tilde{\Omega}.
\end{array}
\nonumber
\end{equation}

Using (\ref{eq:FA_lin_gen_coef_m}), one can easily derive the set of constants (\ref{eq:FA_lin_gen_coef_n}), which define the dynamics of the whole particle.

The obtained expressions for the nanoparticle trajectories let us to write the analytical relation for the power loss $q$.  Direct integration of (\ref{eq:def_q}) with substitution of (\ref{eq:FA_lin_gen_sol}), (\ref{eq:FA_lin_gen_coef_m}) and (\ref{eq:FA_lin_gen_coef_n}) yields the following formula:

\begin{equation}
    \begin{array}{lcl}
    q\!\!&=&\!\! 0.5\tilde{\Omega}\Omega_{r}(b_{m}h\cos\theta_{0}\cos\varphi_{0}-a_{m}\rho h\cos\theta_{0}\sin\varphi_{0}-\\
    \!\!&-&\!\! d_{m}h\sin\varphi_{0}-c_{m}\rho h\cos\varphi_{0}+b_{m}a_{n}-a_{m}b_{n}+d_{m}c_{n}-\\
    \!\!&-&\!\! c_{m}d_{n}).
    \label{eq:FA_q}
    \end{array}
    \end{equation}

The dependence of $q$ on the system parameters, especially on the external field frequency, is of great interest and will be analyzed below. But, the comparison of this result with similar one obtained within other approximations, such as the FP-model and the RD-model, is no less interesting.

\subsection{Oscillations of the Magnetic Moment in a Fixed Nanoparticle}

As the next stage, let us consider the magnetic dynamics only within the FP-model. As in the previous case, the linearized equations are written under the assumption $m_{x'}, m_{y'} \sim h$, and all the nonlinear terms with respect to $h$ are dropped. Using (\ref{eq:h_prime}), we, finally, obtain from (\ref{eq:Red_Eq_FP_m}) the linearized system of equations for $\textbf{m}$ as follows

\begin{equation}
\begin{array}{l}
    \dot{m}_{x'}= - \Omega_{r} \left(\dot{m}_{y'}-h_{y'}\right) - \alpha \dot{m}_{y'},\\
    [2pt]
    \dot{m}_{y'}=  \Omega_{r} \left(\dot{m}_{x'}-h_{x'}\right)- \alpha \dot{m}_{x'}.\\
\end{array}
 \label{eq:Red_Eq_FP_lin}
\end{equation}

Then, the general form of the solution of (\ref{eq:Red_Eq_FP_lin}) can be easily written in the standard form

\begin{equation}
\begin{array}{lcl}
    m_{x'}= a_{fp}\cos \Omega t + b_{fp}\sin \Omega t, \\
    [2pt]
    m_{y'}= c_{fp}\cos \Omega t + d_{fp}\sin \Omega t, \\
    [2pt]
\end{array}
\label{eq:FP_lin_gen_sol}
\end{equation}

where $a_{fp}$, $b_{fp}$, $c_{fp}$, and $d_{fp}$ are the oscillation amplitudes for magnetic the moment inside the immobilized nanoparticle. Substitution of (\ref{eq:FP_lin_gen_sol}) into (\ref{eq:Red_Eq_FP_lin}) yields the system of linear algebraic equations for the desired amplitudes

\begin{equation}
\begin{array}{lcl}
    \tilde{\Omega}a_{fp} \!\!&=&\!\! (d_{fp} - \sigma h \cos \varphi_{0}) - \alpha \tilde{\Omega} c_{fp}, \\
    [2pt]
    \tilde{\Omega}b_{fp} \!\!&=&\!\! - (c_{fp} + h \sin \varphi_{0}) - \alpha \tilde{\Omega} d_{fp}, \\
    [2pt]
    \tilde{\Omega}c_{fp} \!\!&=&\!\! - (b_{fp} - \sigma h \cos \theta_{0} \sin \varphi_{0}) + \alpha \tilde{\Omega} a_{fp}, \\
    [2pt]
    \tilde{\Omega}d_{fp} \!\!&=&\!\! (a_{fp} - h \cos \theta_{0} \cos \varphi_{0}) + \alpha \tilde{\Omega} b_{fp}. \\
\end{array}
\label{eq:FP_alg_gen}
\end{equation}

After the calculations, we find the solution of (\ref{eq:FP_alg_gen})

\begin{equation}
\begin{array}{lcl}
a_{fp} = - Z_{fp}^{-1}\left[\tilde{\Omega}^{fp}_{1} B_{fp} + \tilde{\Omega}^{fp}_{2} A_{fp} \right], \\
[2pt]
b_{fp} = Z_{fp}^{-1}\left[\tilde{\Omega}^{fp}_{1} C_{fp} + \tilde{\Omega}^{fp}_{2} D_{fp} \right], \\
[2pt]
a_{fp} = Z_{fp}^{-1}\left[\tilde{\Omega}^{fp}_{1} A_{fp} - \tilde{\Omega}^{fp}_{2} B_{fp} \right], \\
[2pt]
d_{fp} = Z_{fp}^{-1}\left[-\tilde{\Omega}^{fp}_{1} D_{fp} + \tilde{\Omega}^{fp}_{2} C_{fp} \right], \\
\end{array}
\label{eq:FP_lin_gen_coef_m}
\end{equation}

where

\begin{equation}
    \begin{array}{lcl}
    Z_{fp} = 4\alpha^2 \tilde{\Omega}^{4} + \left((\alpha^{2}-1)\tilde{\Omega}^{-2} + 1\right)^2,
    \label{eq:FA_lin_gen_Z}
    \end{array}
\end{equation}

\begin{equation}
    \begin{array}{lcl}
    \Omega^{fp}_{1}=2\alpha \tilde{\Omega}^{2},\\
    [2pt]
    \Omega^{fp}_{2}=(\alpha^{2}-1)\tilde{\Omega}^{2} + 1,\\
\end{array}
\nonumber
\end{equation}

\begin{equation}
    \begin{array}{lcl}
    A_{fp}= \sigma h \tilde{\Omega} \left(  \alpha\cos\theta_{0}\sin\varphi_0 - \cos\varphi_{0} \right) - h \cos\theta_{0}\cos\varphi_{0},\\
    [2pt]
    B_{fp}=  \sigma h \tilde{\Omega} \left( \cos\varphi_{0}\sin\varphi_0 +  \alpha\cos\varphi_{0} \right) + h \sin\varphi_{0},\\
    [2pt]
    C_{fp}=  h \tilde{\Omega}  \left(\cos\theta_{0}\cos\varphi_{0} - \alpha \sin\varphi_0 \right) + \sigma h \cos\varphi_{0},\\
    [2pt]
    D_{fp}=  h \tilde{\Omega}  \left(\alpha \cos\theta_{0}\cos\varphi_{0} + \sin\varphi_0 \right) + \sigma h \cos\theta_{0}\sin\varphi_{0}.\\
\end{array}
\nonumber
\end{equation}

The power loss in this case can be also found by direct integration of (\ref{eq:def_q}) with substitution of (\ref{eq:FP_lin_gen_sol}) and (\ref{eq:FP_lin_gen_coef_m})

\begin{equation}
\begin{array}{lcl}
    q\!\!&=&\!\! 0.5 h \tilde{\Omega} \Omega_{r} Z_{fp}^{-1} \{ \tilde{\Omega}^{fp}_{1}  [2\sigma h \cos\theta_{0} + \\
    [4pt]
    \!\!&+&\!\! h \tilde{\Omega} D ] + \tilde{\Omega}^{fp}_{2} \alpha h \tilde{\Omega}  D \},
    \label{eq:FP_q}
\end{array}
\end{equation}

where

\begin{equation}
    D =  \cos^2 \theta_0(\cos^2 \varphi_0 + \sigma^2\sin^2 \varphi_0) + \sigma^2\cos^2 \varphi_0 + \sin^2 \varphi_0.
    \label{eq:D}
\end{equation}

The obtained expression (\ref{eq:FP_q}) is similar to the reported in \cite{PhysRevB.91.054425}, but accounts an arbitrary orientation of the nanoparticle easy axis. Despite the quantitative difference caused by the turn of the easy axis, the qualitative character of the frequency behavior of $q$ remains.

\subsection{Oscillations of a Nanoparticle with the Locked Magnetic Moment}

And finally, we consider within the same framework the widely used approach when the nanoparticle magnetic moment is rigidly bound with the nanoparticle crystal lattice. In this so-called RD-model, the linearized equations have the simplest form. Expanding the vector equation (\ref{eq:Red_Eq_FM_n}) and taking into account (\ref{eq:h_prime}), we  write the linearized system of equations for $\textbf{n}$ as follows

\begin{equation}
\begin{array}{l}
    \dot{n}_{x'}= \Omega_{cr} h_{x'},\\
    [2pt]
    \dot{n}_{y'}= \Omega_{cr} h_{y'}.\\
\end{array}
 \label{eq:Red_Eq_FM_lin}
\end{equation}

As in the previous case, we use the trigonometric representation of the solution of (\ref{eq:Red_Eq_FM_lin})

\begin{equation}
\begin{array}{lcl}
    n_{x'}= a_{rd}\cos \Omega t + b_{rd}\sin \Omega t, \\
    [2pt]
    n_{y'}= c_{rd}\cos \Omega t + d_{rd}\sin \Omega t. \\
    [2pt]
\end{array}
\label{eq:FM_lin_gen_sol}
\end{equation}

After direct substitution of (\ref{eq:FM_lin_gen_sol}) into (\ref{eq:Red_Eq_FM_lin}), one can easily obtain the unknown constants, which are the amplitudes of the vector $\mathbf{n}$

\begin{equation}
\begin{array}{lcl}
    a_{rd} \!\!&=&\!\! h \Omega_{cr} \sin \varphi_{0}/ \Omega, \\
    [2pt]
    b_{rd} \!\!&=&\!\! h \Omega_{cr} \cos \theta_{0} \cos \varphi_{0}/ \Omega, \\
    [2pt]
    c_{rd} \!\!&=&\!\! - h \Omega_{cr} \cos \varphi_{0}/ \Omega, \\
    [2pt]
    d_{rd} \!\!&=&\!\! h \Omega_{cr} \cos \theta_{0} \sin \varphi_{0}/ \Omega. \\
\end{array}
\label{eq:FM_lin_gen_coef_n}
\end{equation}

And at last, we can directly find the power loss from (\ref{eq:def_q}) substituting (\ref{eq:FM_lin_gen_sol}) and (\ref{eq:FM_lin_gen_coef_n})

\begin{equation}
\begin{array}{lcl}
    q\!\!&=&\!\! 0.5 \Omega_{cr} h^2 D.
    \label{eq:FM_q}
\end{array}
\end{equation}

It is remarkable that $q$ does not depend on the frequency, because while $\Omega$ increases, the coefficients (\ref{eq:FM_lin_gen_coef_n}) decrease proportionally that compensates the possible growth of the power loss.

\section{Discussion and Conclusions}

We have considered the response of a uniaxial ferromagnetic nanoparticle placed into a viscous fluid to an alternating field in the linear approximation for three models, namely, the FA-model (viscously coupled nanoparticle with a finite anisotropy), the FP-model (fixed particle), and the RD-model (rigid dipole). As a result, we have obtained the expressions for the nanoparticle trajectories and for the power loss produced by both the rotation of a nanoparticle in a viscous media and the internal damping precession of the nanoparticles magnetic moment. Our main aims were the understanding of 1) the power loss behavior depending on different parameters; 2) the role of dissipation mechanisms when they both are present; and 3) the correlations between the mechanical rotation of a nanoparticle and the internal motion of its magnetic moment. The analysis of three approximations simultaneously helps us comprehend the restrictions and applicability limits that, in turn, allows to systematize the results obtained by other authors. The relevance of our findings is closely bounded with the application issues, such as heating rate during magnetic fluid hyperthermia or absorption frequency range of the microwave absorbing materials.

The comparison of the expressions for the power loss derived in the previous section allows a number of conclusions, and some of them are rather unexpected at first sight.  Firstly, the role of the internal magnetic motion is primary. As follows from (\ref{eq:FA_q}) and (\ref{eq:FP_q}), the dependencies of the dimensionless power loss on the reduced frequency $q(\tilde{\Omega})$ for the model of fixed particle and for the model of viscously coupled nanoparticle with a finite anisotropy are similar: they both demonstrate a resonant behavior. At the same time, for the model of rigid dipole it remains constant (see (\ref{eq:FM_q})). Therefore, the dynamics of the magnetic moment represented by unit vector $\mathbf{m}$ determines the resulting power loss in a wide range of realistic parameters. But the quantitative comparison of these dependencies lets us assume that the easy axis oscillations can considerably modify the power loss induced by damping precession. The reasons for that are in the character of the collective motion of easy axis, which is represented by vector $\mathbf{n}$, and magnetic moment $\mathbf{m}$. Although the harmonic motion takes place, the ratio of their phases and amplitudes can lead to quite different values of the energy dissipation in the system. Further we consider the behavior of $q(\tilde{\Omega})$ in context of the features of the $\mathbf{m}$ and $\mathbf{n}$ motion.

\begin{figure}
    \centering
    \includegraphics [width=0.95\linewidth]{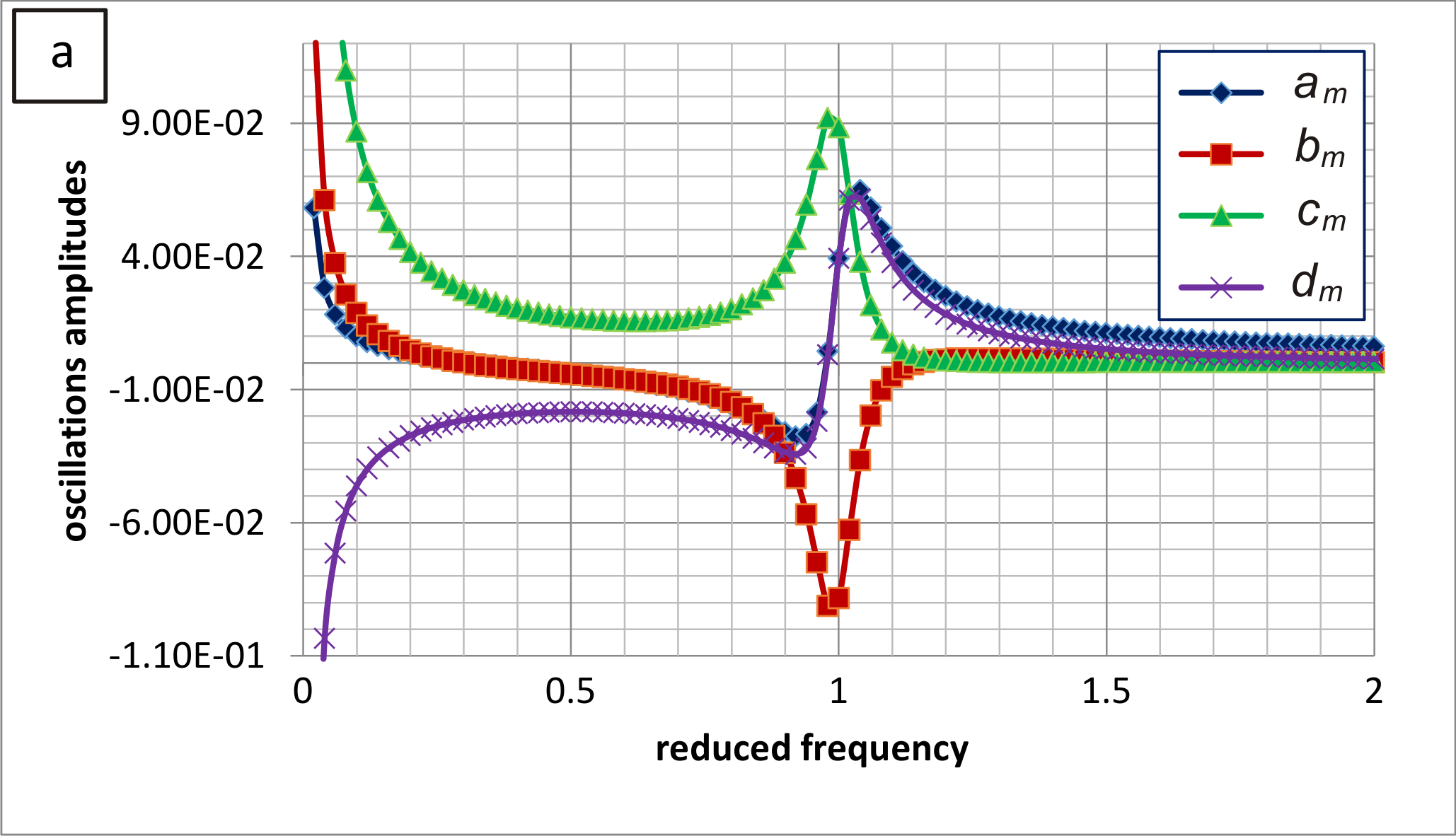}
    \includegraphics [width=0.95\linewidth]{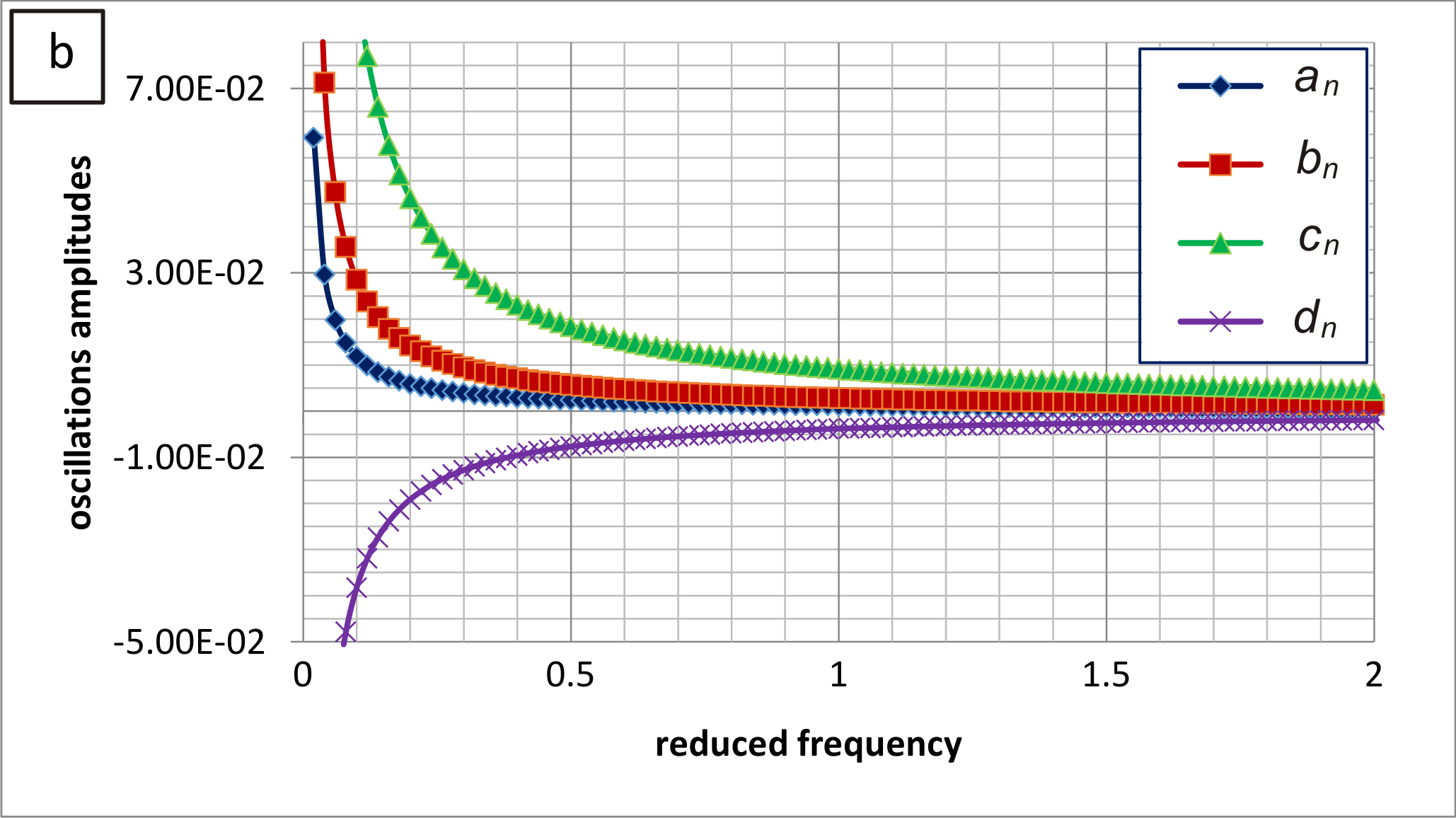}
    \caption {\label{fig:Coef_1} (Color online) The dependencies of the amplitudes of coupled oscillations of the magnetic moment (\ref{eq:FA_lin_gen_coef_m}) and the easy axis (\ref{eq:FA_lin_gen_coef_n}) on the field frequency. The parameters used are $M = 338~\textrm{G}$, $H_a = 910~\textrm{Oe}$, $\eta = 0.006~\textrm{P}$, $\alpha = 0.05$ that corresponds to maghemite nanoparticles ($\gamma-\textrm{Fe}_{2}\textrm{O}_{3}$) in water at the temperature of $42~^{\circ}\textrm{C}$, $\sigma = - 1$, $h = 0.01$, $\theta_0 = 0.4 \pi$, $\varphi_0 = 0.125 \pi$.}
\end{figure}

\begin{figure}
    \centering
    \includegraphics [width=0.95\linewidth]{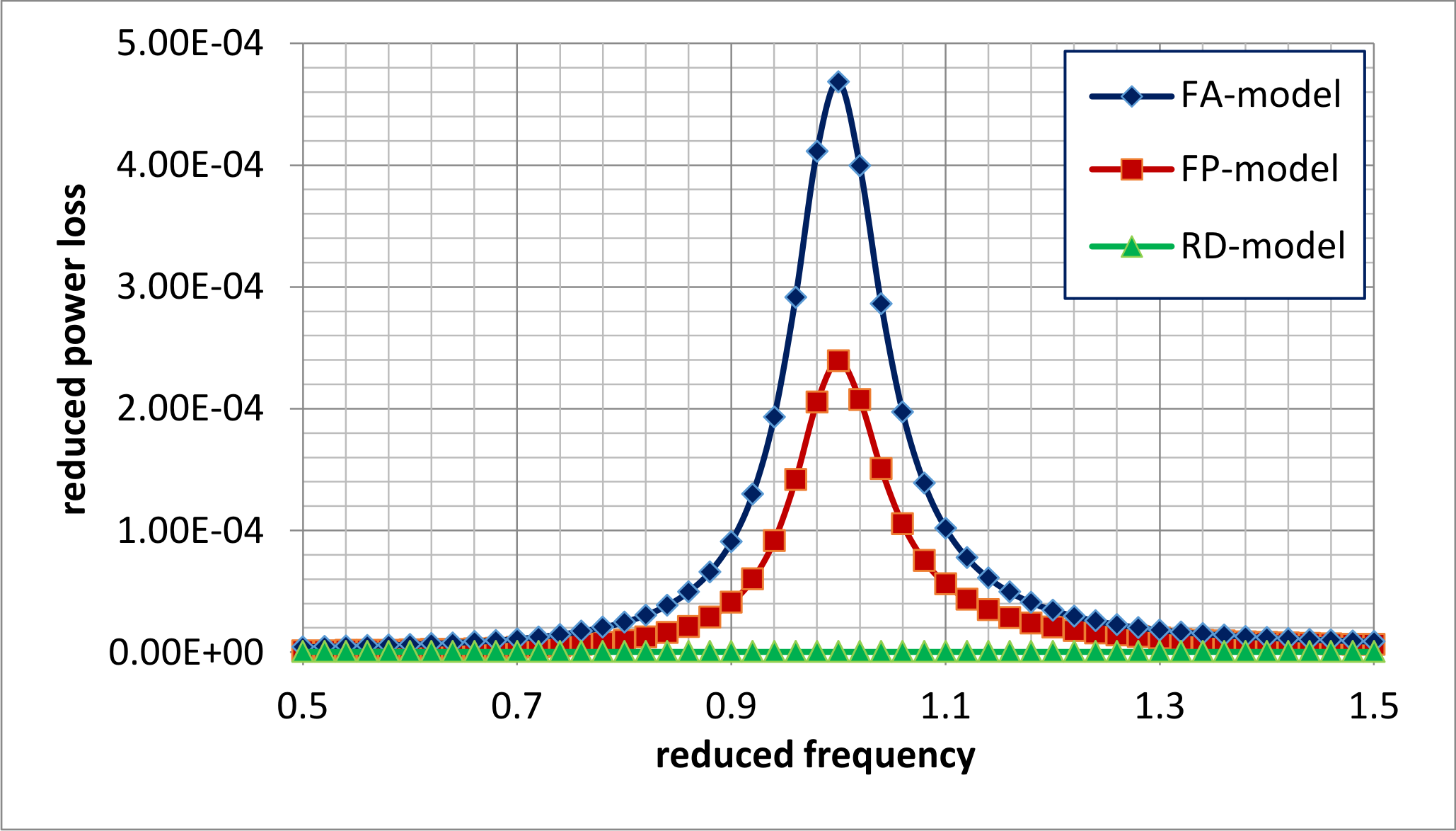}
    \caption {\label{fig:Power_loss_1} (Color online) The frequency dependencies of the power loss for the cases of rigid dipole (RD-model), fixed particle (FP-model), and viscously coupled nanoparticle with a finite anisotropy (FA-model). The parameters used are the same as in the caption to Fig.~\ref{fig:Coef_1}.}
\end{figure}

The behavior of $q(\tilde{\Omega})$ is caused by the features of coefficients $a_m(\tilde{\Omega})$, $b_m(\tilde{\Omega})$, $c_m(\tilde{\Omega})$, and $d_m(\tilde{\Omega})$, which determine the $\mathbf{m}$ dynamics, and $a_n(\tilde{\Omega})$, $b_n(\tilde{\Omega})$, $c_n(\tilde{\Omega})$, and $d_n(\tilde{\Omega})$ defining the dynamics of $\mathbf{n}$ (see Fig.~\ref{fig:Coef_1}). As seen, for frequencies far from the resonance one, vectors $\mathbf{n}$ and $\mathbf{m}$ almost coincide and are rotated synchronously. Here, the model of viscously coupled nanoparticle with a finite anisotropy and the model of fixed particle yield very close values of the power loss. But near the resonance, in the vicinity of $\tilde{\Omega} = 1$, coefficients $a_m(\tilde{\Omega})$, $b_m(\tilde{\Omega})$, $c_m(\tilde{\Omega})$, and $d_m(\tilde{\Omega})$ have the pronounced maxima and change the signs, while coefficients $a_n(\tilde{\Omega})$, $b_n(\tilde{\Omega})$, $c_n(\tilde{\Omega})$, and $d_n(\tilde{\Omega})$ remain the same. Therefore, vectors $\mathbf{n}$ and $\mathbf{m}$ are rotated in an asynchronous way now that leads to a larger angle between the magnetic moment and the resulting or effective field $\mathbf{h}_{eff}$. Together with increasing precession angle of $\mathbf{m}$, this causes the growth of the power loss compared with the case of fixed particle (see Fig.~\ref{fig:Power_loss_1}).

\begin{figure}
    \centering
    \includegraphics [width=0.95\linewidth]{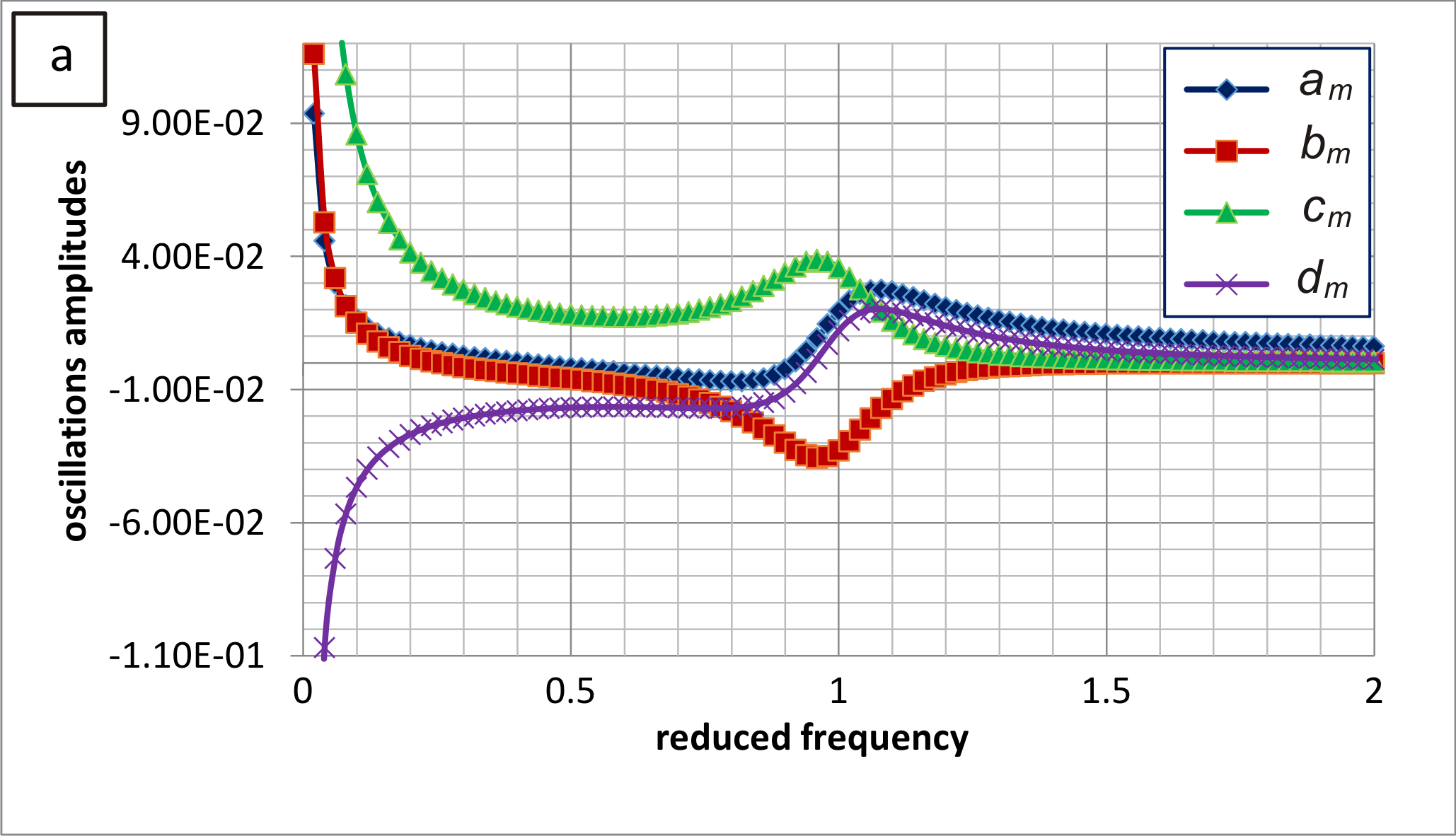}
    \includegraphics [width=0.95\linewidth]{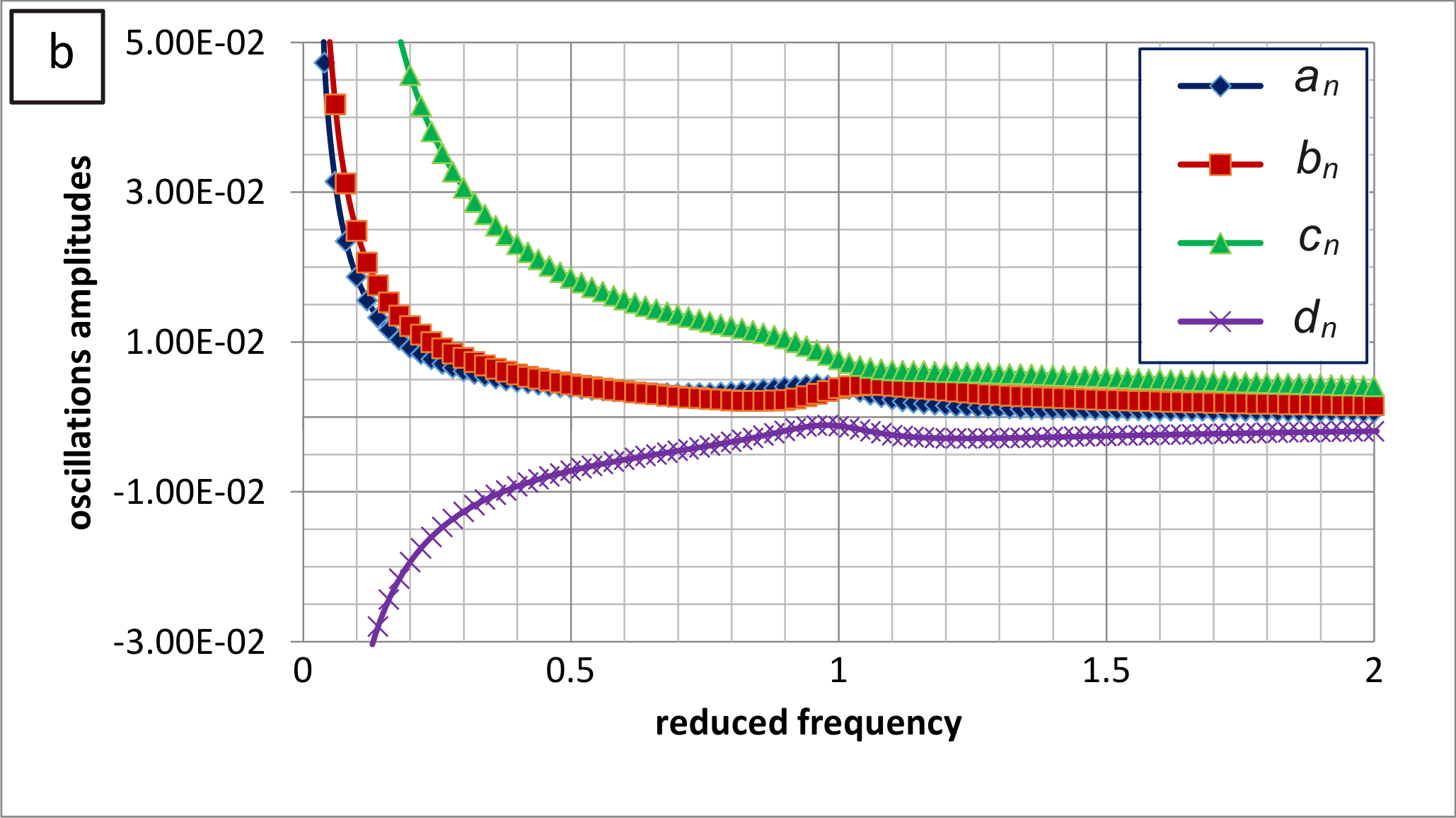}
    \caption {\label{fig:Coef_2} (Color online) The dependencies of the amplitudes of coupled oscillations of the magnetic moment (\ref{eq:FA_lin_gen_coef_m}) and the easy axis (\ref{eq:FA_lin_gen_coef_n}) on the field frequency. The parameters used are the same as in the caption to Fig.~\ref{fig:Coef_1}, but $\eta = 4.0^{-5}~\textrm{P}$.}
\end{figure}

\begin{figure}
    \centering
    \includegraphics [width=0.95\linewidth]{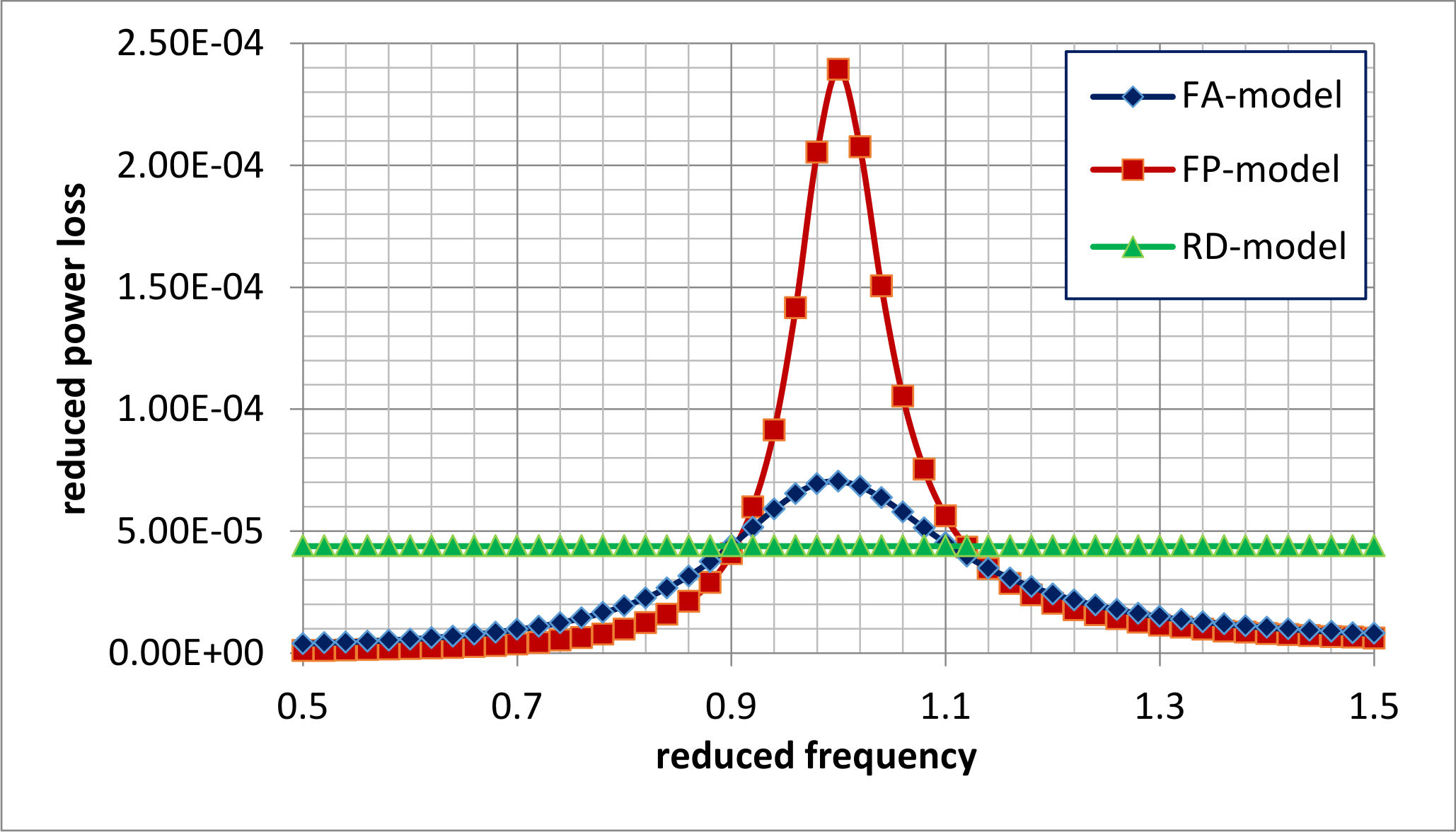}
    \caption {\label{fig:Power_loss_2} (Color online) The frequency dependencies of the power loss for the cases of rigid dipole (RD-model), fixed particle (FP-model), and viscously coupled nanoparticle with a finite anisotropy (FA-model). The parameters used are the same as in the caption to Fig.~\ref{fig:Coef_1}, but $\eta = 4.0^{-5}~\textrm{P}$.}
\end{figure}

For small viscosity, vector $\mathbf{n}$ becomes more susceptible to the external field, and the rotating magnetic moment can easily involve a whole nanoparticle into rotation. But this does not induce a more intense motion in result. Firstly, a considerable decrease in the coefficients $a_m(\tilde{\Omega})$, $b_m(\tilde{\Omega})$, $c_m(\tilde{\Omega})$, and $d_m(\tilde{\Omega})$ near the resonance takes place in comparison with the case of larger viscosity. Then, only $b_m(\tilde{\Omega})$ and $d_m(\tilde{\Omega})$ change the signs now (see Fig.~\ref{fig:Coef_2}). Finally, the dependencies $a_n(\tilde{\Omega})$, $b_n(\tilde{\Omega})$, $c_n(\tilde{\Omega})$, and $d_n(\tilde{\Omega})$ get the local maxima (Fig.~\ref{fig:Coef_2}) and slightly decrease in absolute values in the vicinity of $\tilde{\Omega} = 1$. Therefore, the effect of the pronounced asynchronous rotation of $\mathbf{n}$ and $\mathbf{m}$, which is actual for the foregoing case, eliminates now, and they become almost parallel for a whole range of frequencies. Since the angle between the magnetic moment and the resulting field is reduced, the model of viscously coupled nanoparticle with a finite anisotropy predicts lower values of the power loss than the model of fixed particle near the resonance (Fig.~\ref{fig:Power_loss_2}).

The situation described above is an origin for extreme sensitivity of the power loss to the system parameters, which may be useful in the applications and can be utilized in a number of cases. In contrary, in other cases such sensitivity can be very undesirable, and we have to take measures to prevent it. Independently of the further purposes, one needs to investigate the influence of the main parameters in detail. It is especially important for the design of the nanoparticle ensembles with the specified properties for key applications, such as microwave absorbing or magnetic fluid hyperthermia, where the heating or/and absorbing rates are the primary characteristics.

\begin{figure}
    \centering
    \includegraphics [width=0.95\linewidth]{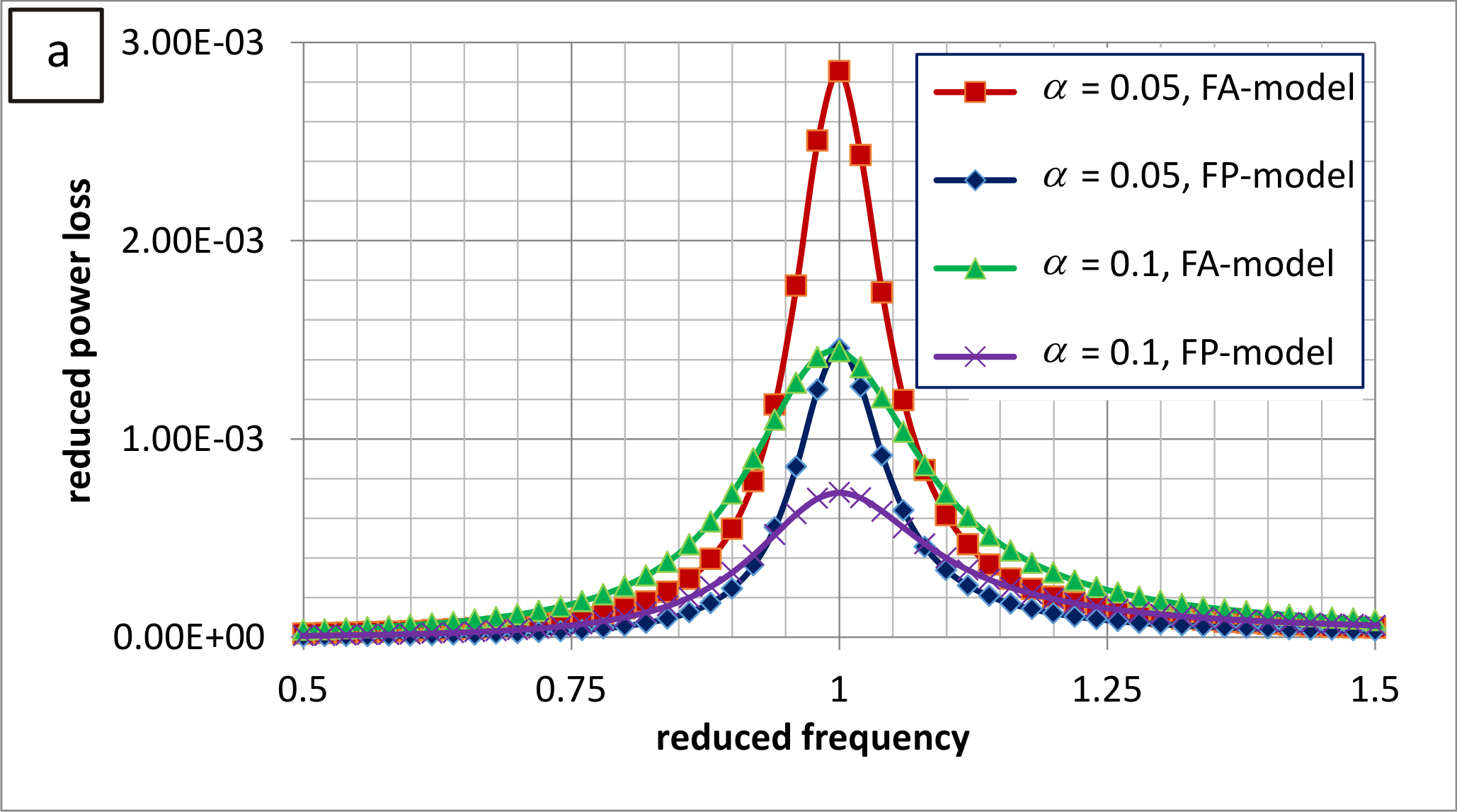}
    \includegraphics [width=0.95\linewidth]{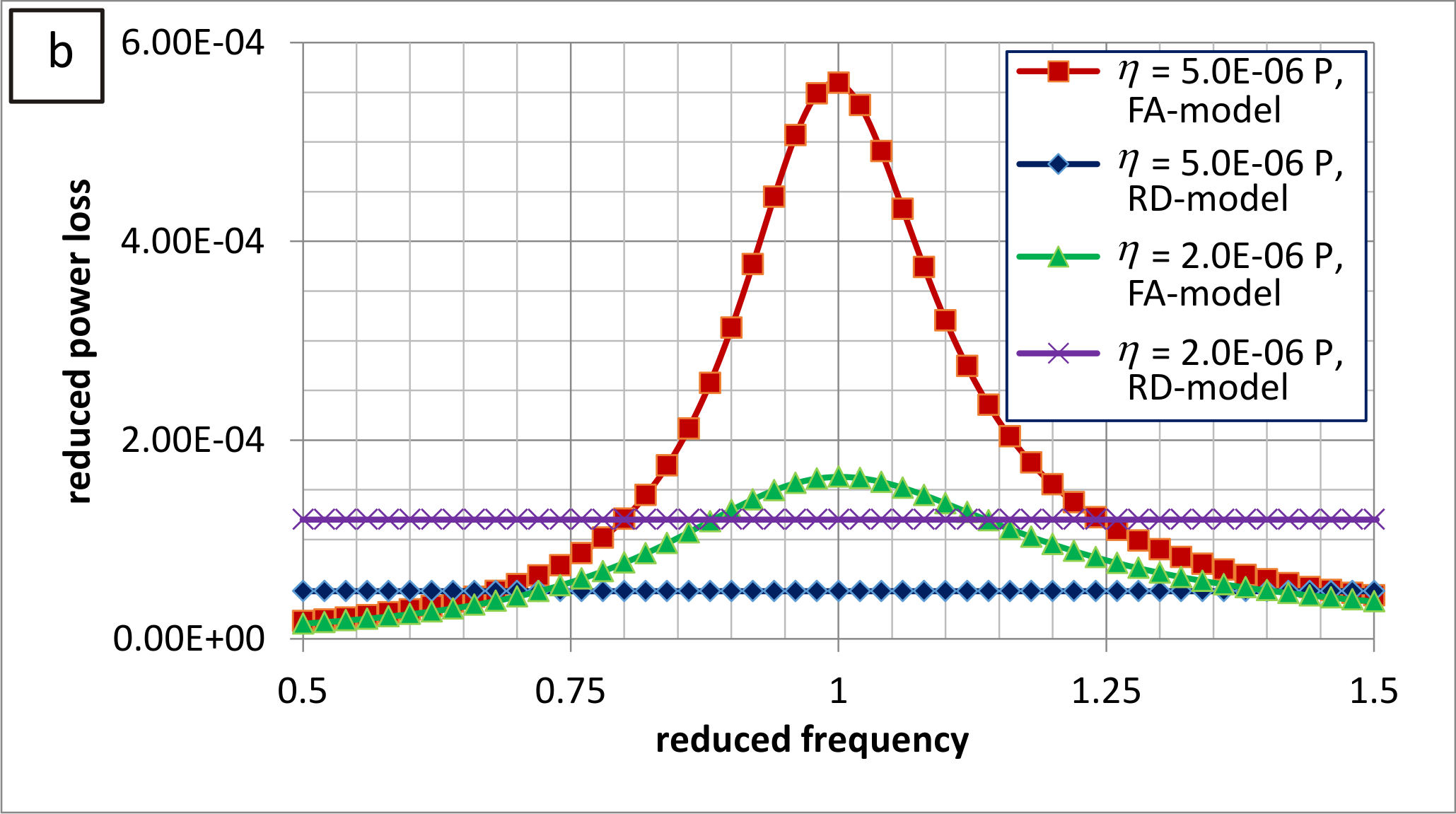}
    \caption {\label{fig:Pl_alpha_eta} (Color online) The sensitivity of the power loss to the attenuation parameters. Plot (a): fixed particle (FP-model) and viscously coupled nanoparticle with a finite anisotropy (FA-model) and different values of the magnetic damping parameter $\alpha$. Plot (b): rigid dipole (RD-model) and viscously coupled nanoparticle with a finite anisotropy (FA-model) and different values of the $\eta$ and results obtained for the cases rigid dipole (RD-model) and viscously coupled nanoparticle with a finite anisotropy (FA-model). The parameters used here and not stated in the figure legend are the same as in the caption to Fig.~\ref{fig:Coef_1}, but $\theta_0 = 0.25 \pi$.}
\end{figure}

In this regard, the similar parameters $\alpha$ and $\eta$ are the most interesting. In Fig.~\ref{fig:Pl_alpha_eta}a,  the comparison of the power loss for two values of $\alpha$ are plotted using the fixed particle approximation and the approximation of viscously coupled nanoparticle with a finite anisotropy. As expected, the decrease in $\alpha$ leads to the proportional increase in the power loss for both approximations. At the same time, the change in $\eta$ results in different behavior of the power loss obtained using the rigid dipole approximation and the approximation of viscously coupled nanoparticle with a finite anisotropy (see Fig.~\ref{fig:Pl_alpha_eta}b). For the first case, the proportional growth of $q(\tilde{\Omega})$ with decreasing $\eta$ takes place. But for the second case, account of the finite anisotropy leads to the opposite results. Here we report a nonlinear growth in $q(\tilde{\Omega})$ with increasing viscosity $\eta$. As it was explained above, the origin of this effect lies in the relative motion of vectors $\mathbf{n}$ and $\mathbf{m}$. Then, to estimate the applicability of the model of rigid dipole, one needs to compare the power loss values for these two cases. As seen from Fig.~\ref{fig:Pl_alpha_eta}b, various situations are possible because there are two different behavior types when $\mathbf{m}$ is unlocked. The first type is the asynchronous oscillations of $\mathbf{m}$ and $\mathbf{n}$, wherein the values of $q(\tilde{\Omega})$ for the model of viscously coupled nanoparticle with a finite anisotropy can be considerably larger than the values predicted by the model of rigid dipole. The second type is the synchronous motion of the magnetic moment and the easy axis. Here, both dissipation mechanisms are suppressed because the amplitudes of $\mathbf{n}$ and $\mathbf{m}$ oscillations become smaller. As a result, the power loss for the finite anisotropy case can be substantially lower than the value obtained for the model of rigid dipole. This allows us to conclude about a low applicability of the model of rigid dipole in a high frequency limit.

\begin{figure}
    \centering
    \includegraphics [width=0.95\linewidth]{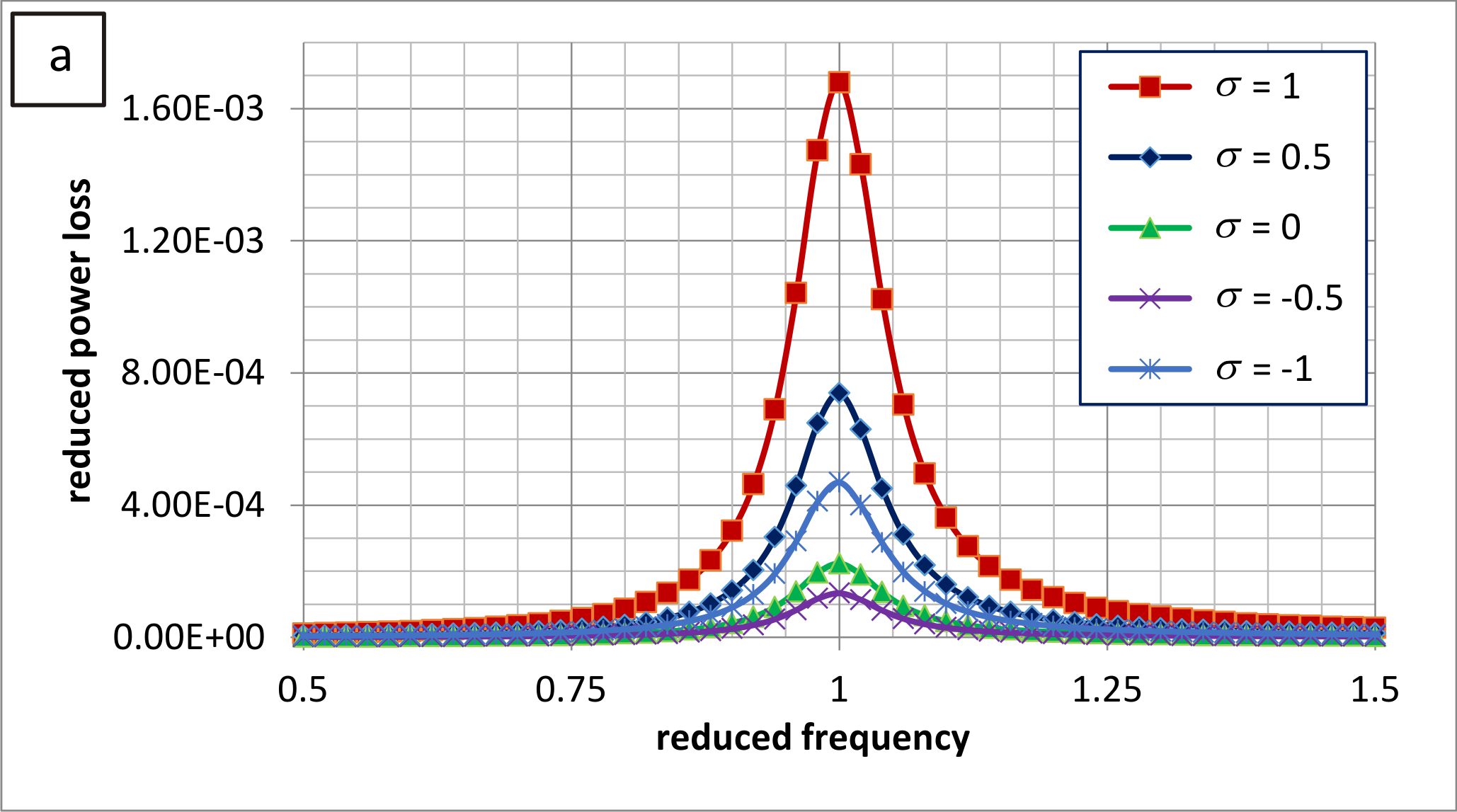}
    \includegraphics [width=0.95\linewidth]{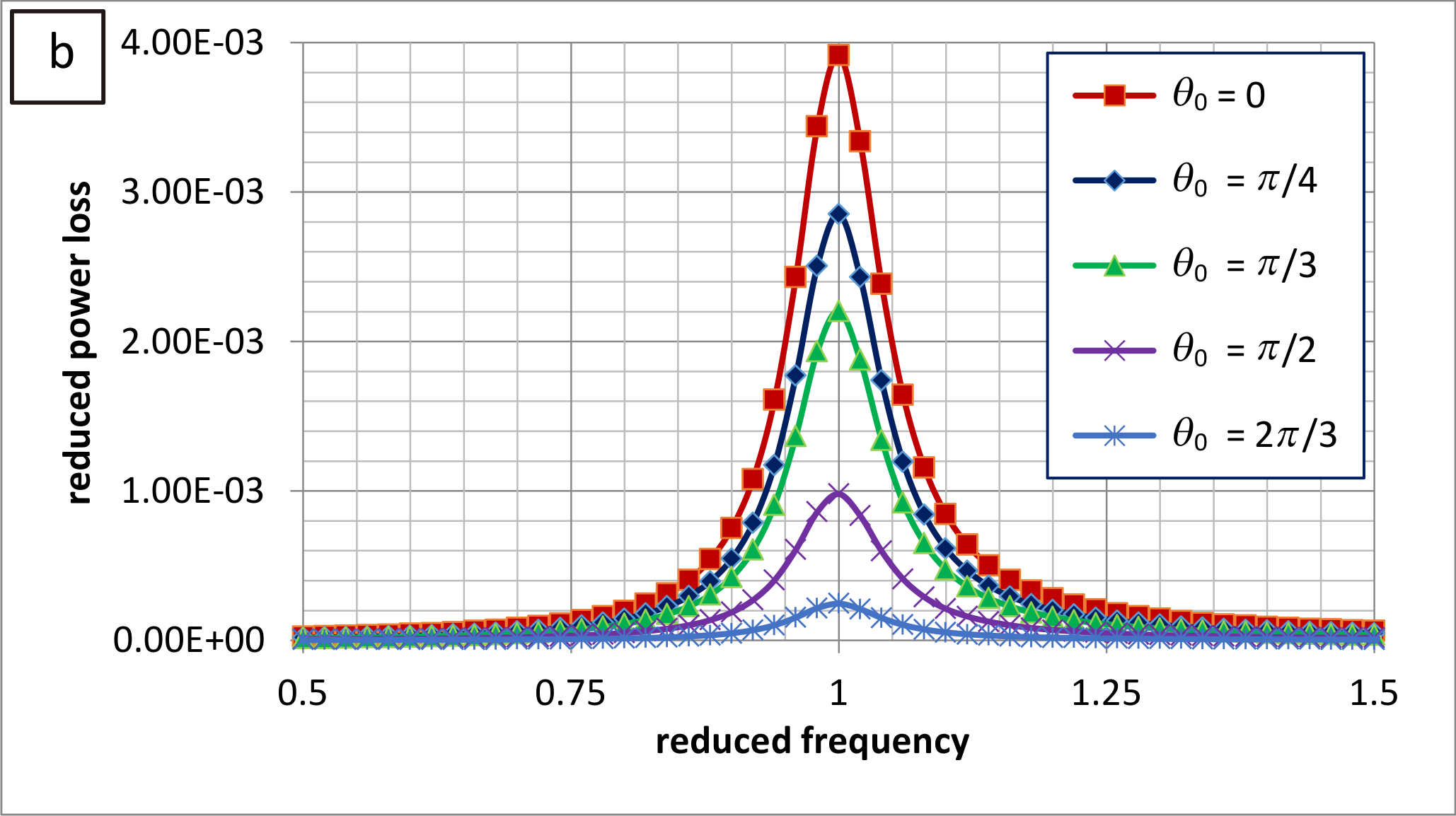}
    \caption {\label{fig:Pl_sigma_theta} (Color online) The sensitivity of the power loss to the orientation of the nanoparticle with respect to the external field for the case of viscously coupled nanoparticle with a finite anisotropy (FA-model). The parameters used here and not stated in the figure legend are the same as in the caption to Fig.~\ref{fig:Coef_1}, but $\theta_0 = 0.25 \pi$ for the plot (a) and  $\sigma = 1$ for the plot (b).}
\end{figure}

Another important issue which needs to be accounted is the influence of the external field orientation with respect to the nanoparticle position. As follows from (\ref{eq:FA_q}), (\ref{eq:FP_q}), (\ref{eq:FM_q}), this orientation is defined by the polarization type and the initial position of the easy axis. The model of rigid dipole predicts the difference of the power loss not more than two times when $\sigma$ varies in the range of $[-1...1]$. In accordance with two other models, the dependence of the power loss on $\sigma$ is more strong. As seen from Fig.~\ref{fig:Pl_sigma_theta}a, $q(\tilde{\Omega})$ can be at least $10$ times different depending on $\sigma$ for the model of viscously coupled nanoparticle with a finite anisotropy. Here we need to note that this dependence is not linear and the lowest curve $q(\tilde{\Omega})$ does not correspond to $\sigma = 0$ or $\sigma=\pm 1$. The initial position of the easy axis given by angle $\theta_0$ essentially influences the power loss as well. As seen from Fig.~\ref{fig:Pl_sigma_theta}b, this difference may be at least 20 times. Since nanoparticles in real ferrofluids are non-uniformly distributed, one can highlight the following. Firstly, the dipole interaction, which tries to arrange the ensemble, can considerably influence the power loss. Secondly, an external magnetic field gradient, which is used for the ferrofluid control during hyperthermia, also defines the power loss. And, thirdly, we can easily control the power loss in a wide range of values by a permanent external field, which specifies the direction of the nanoparticle easy axis.

We summarize our findings as follows. 1) The small oscillations mode is considered for the coupled magnetic and mechanical motion for the viscously coupled nanoparticle with a finite anisotropy. This mode takes place when the amplitude of the external alternating field is much smaller than the value of the nanoparticle uniaxial anisotropy field ($H \ll H_a$). 2) The damping precession of the magnetic moment inside the nanoparticle primarily determines the value of the power loss and the resonance character of its frequency dependence. 3) The power loss can be significantly changed by the nanoparticle easy axis motion. For the realistic system parameters, the power loss obtained for the model of viscously coupled nanoparticle with a finite anisotropy is larger than the value obtained for the fixed particle model. 4) The decrease in the fluid carrier viscosity leads to the nonproportional decrease in the power loss, which near the resonance can be much smaller than the value obtained for the fixed particle model. Such complicated correlation between the magnetic dynamics and the mechanical motion does not allow to separate the contributions of these two mechanisms into dissipation. 5) The power loss is extremely sensitive to the system parameters and the nanoparticle initial position. It should be taken into account and can be used, for example, for the control of the heating and absorbing rates. Although the results are obtained in the dynamical approach, they establish the limitation for more precise models which account thermal fluctuations and inter-particle interaction.

\section*{Acknowledgements}
The authors are grateful to the Ministry of Education and Science of Ukraine for partial financial support under Grant No.~0116U002622.

\section*{References}

\bibliography{Lyutyy_Sumy_Ukraine_FNP_in_Fluid_Revised}

\begin{thebibliography}{10}
\expandafter\ifx\csname url\endcsname\relax
  \def\url#1{\texttt{#1}}\fi
\expandafter\ifx\csname urlprefix\endcsname\relax\def\urlprefix{URL }\fi
\expandafter\ifx\csname href\endcsname\relax
  \def\href#1#2{#2} \def\path#1{#1}\fi

\bibitem{Rosensweig2002370}
R.~Rosensweig,
  \href{http://www.sciencedirect.com/science/article/pii/S0304885302007060}{Heating
  magnetic fluid with alternating magnetic field}, Journal of Magnetism and
  Magnetic Materials 252 (2002) 370 -- 374, proceedings of the 9th
  International Conference on Magnetic Fluids.
\newblock \href
  {http://dx.doi.org/http://dx.doi.org/10.1016/S0304-8853(02)00706-0}
  {\path{doi:http://dx.doi.org/10.1016/S0304-8853(02)00706-0}}.
\newline\urlprefix\url{http://www.sciencedirect.com/science/article/pii/S0304885302007060}

\bibitem{andr2006magnetism}
W.~Andr\"{a}, H.~Nowak,
  \href{http://onlinelibrary.wiley.com/book/10.1002/9783527610174}{Magnetism in
  Medicine: A Handbook}.
\newblock \href {http://dx.doi.org/10.1002/9783527610174}
  {\path{doi:10.1002/9783527610174}}.
\newline\urlprefix\url{http://onlinelibrary.wiley.com/book/10.1002/9783527610174}

\bibitem{0038-5670-17-2-R02}
M.~I. Shliomis, \href{http://stacks.iop.org/0038-5670/17/i=2/a=R02}{Magnetic
  fluids}, Soviet Physics Uspekhi 17~(2) (1974) 153.
\newline\urlprefix\url{http://stacks.iop.org/0038-5670/17/i=2/a=R02}

\bibitem{PhysRev.130.1677}
W.~F. Brown, \href{https://link.aps.org/doi/10.1103/PhysRev.130.1677}{Thermal
  fluctuations of a single-domain particle}, Phys. Rev. 130 (1963) 1677--1686.
\newblock \href {http://dx.doi.org/10.1103/PhysRev.130.1677}
  {\path{doi:10.1103/PhysRev.130.1677}}.
\newline\urlprefix\url{https://link.aps.org/doi/10.1103/PhysRev.130.1677}

\bibitem{Jordan1999413}
A.~Jordan, R.~Scholz, P.~Wust, H.~Fähling, R.~Felix,
  \href{http://www.sciencedirect.com/science/article/pii/S0304885399000888}{Magnetic
  fluid hyperthermia (mfh): Cancer treatment with \{AC\} magnetic field induced
  excitation of biocompatible superparamagnetic nanoparticles}, Journal of
  Magnetism and Magnetic Materials 201~(1-3) (1999) 413 -- 419.
\newblock \href
  {http://dx.doi.org/http://dx.doi.org/10.1016/S0304-8853(99)00088-8}
  {\path{doi:http://dx.doi.org/10.1016/S0304-8853(99)00088-8}}.
\newline\urlprefix\url{http://www.sciencedirect.com/science/article/pii/S0304885399000888}

\bibitem{0022-3727-36-13-201}
Q.~A. Pankhurst, J.~Connolly, S.~K. Jones, J.~Dobson,
  \href{http://stacks.iop.org/0022-3727/36/i=13/a=201}{Applications of magnetic
  nanoparticles in biomedicine}, Journal of Physics D: Applied Physics 36~(13)
  (2003) R167.
\newline\urlprefix\url{http://stacks.iop.org/0022-3727/36/i=13/a=201}

\bibitem{Raikher2011}
Y.~L. Raikher, V.~I. Stepanov,
  \href{http://dx.doi.org/10.1134/S1063776110061160}{Energy absorption by a
  magnetic nanoparticle suspension in a rotating field}, Journal of
  Experimental and Theoretical Physics 112~(1) (2011) 173--177.
\newblock \href {http://dx.doi.org/10.1134/S1063776110061160}
  {\path{doi:10.1134/S1063776110061160}}.
\newline\urlprefix\url{http://dx.doi.org/10.1134/S1063776110061160}

\bibitem{PhysRevE.83.021401}
Y.~L. Raikher, V.~I. Stepanov,
  \href{https://link.aps.org/doi/10.1103/PhysRevE.83.021401}{Power losses in a
  suspension of magnetic dipoles under a rotating field}, Phys. Rev. E 83
  (2011) 021401.
\newblock \href {http://dx.doi.org/10.1103/PhysRevE.83.021401}
  {\path{doi:10.1103/PhysRevE.83.021401}}.
\newline\urlprefix\url{https://link.aps.org/doi/10.1103/PhysRevE.83.021401}

\bibitem{7753812}
V.~V. Reva, T.~V. Lyutyy, Microwave absorption by a rigid dipole in a viscous
  fluid, in: 2016 II International Young Scientists Forum on Applied Physics
  and Engineering (YSF), 2016, pp. 104--107.
\newblock \href {http://dx.doi.org/10.1109/YSF.2016.7753812}
  {\path{doi:10.1109/YSF.2016.7753812}}.

\bibitem{0953-8984-15-23-313}
B.~U. Felderhof, R.~B. Jones,
  \href{http://stacks.iop.org/0953-8984/15/i=23/a=313}{Mean field theory of the
  nonlinear response of an interacting dipolar system with rotational diffusion
  to an oscillating field}, Journal of Physics: Condensed Matter 15~(23) (2003)
  4011.
\newline\urlprefix\url{http://stacks.iop.org/0953-8984/15/i=23/a=313}

\bibitem{PhysRevE.92.042312}
T.~V. Lyutyy, S.~I. Denisov, V.~V. Reva, Y.~S. Bystrik,
  \href{http://link.aps.org/doi/10.1103/PhysRevE.92.042312}{Rotational
  properties of ferromagnetic nanoparticles driven by a precessing magnetic
  field in a viscous fluid}, Phys. Rev. E 92 (2015) 042312.
\newblock \href {http://dx.doi.org/10.1103/PhysRevE.92.042312}
  {\path{doi:10.1103/PhysRevE.92.042312}}.
\newline\urlprefix\url{http://link.aps.org/doi/10.1103/PhysRevE.92.042312}

\bibitem{PhysRevE.86.061404}
I.~N\'andori, J.~R\'acz,
  \href{http://link.aps.org/doi/10.1103/PhysRevE.86.061404}{Magnetic particle
  hyperthermia: Power losses under circularly polarized field in anisotropic
  nanoparticles}, Phys. Rev. E 86 (2012) 061404.
\newblock \href {http://dx.doi.org/10.1103/PhysRevE.86.061404}
  {\path{doi:10.1103/PhysRevE.86.061404}}.
\newline\urlprefix\url{http://link.aps.org/doi/10.1103/PhysRevE.86.061404}

\bibitem{PhysRevE.93.012607}
J.~R\'acz, P.~F. de~Ch\^atel, I.~A. Szab\'o, L.~Szunyogh, I.~N\'andori,
  \href{http://link.aps.org/doi/10.1103/PhysRevE.93.012607}{Improved efficiency
  of heat generation in nonlinear dynamics of magnetic nanoparticles}, Phys.
  Rev. E 93 (2016) 012607.
\newblock \href {http://dx.doi.org/10.1103/PhysRevE.93.012607}
  {\path{doi:10.1103/PhysRevE.93.012607}}.
\newline\urlprefix\url{http://link.aps.org/doi/10.1103/PhysRevE.93.012607}

\bibitem{PhysRevB.91.054425}
T.~V. Lyutyy, S.~I. Denisov, A.~Y. Peletskyi, C.~Binns,
  \href{http://link.aps.org/doi/10.1103/PhysRevB.91.054425}{Energy dissipation
  in single-domain ferromagnetic nanoparticles: Dynamical approach}, Phys. Rev.
  B 91 (2015) 054425.
\newblock \href {http://dx.doi.org/10.1103/PhysRevB.91.054425}
  {\path{doi:10.1103/PhysRevB.91.054425}}.
\newline\urlprefix\url{http://link.aps.org/doi/10.1103/PhysRevB.91.054425}

\bibitem{PhysRevB.85.045435}
C.~Haase, U.~Nowak,
  \href{https://link.aps.org/doi/10.1103/PhysRevB.85.045435}{Role of
  dipole-dipole interactions for hyperthermia heating of magnetic nanoparticle
  ensembles}, Phys. Rev. B 85 (2012) 045435.
\newblock \href {http://dx.doi.org/10.1103/PhysRevB.85.045435}
  {\path{doi:10.1103/PhysRevB.85.045435}}.
\newline\urlprefix\url{https://link.aps.org/doi/10.1103/PhysRevB.85.045435}

\bibitem{PhysRevB.89.014403}
G.~T. Landi, \href{https://link.aps.org/doi/10.1103/PhysRevB.89.014403}{Role of
  dipolar interaction in magnetic hyperthermia}, Phys. Rev. B 89 (2014) 014403.
\newblock \href {http://dx.doi.org/10.1103/PhysRevB.89.014403}
  {\path{doi:10.1103/PhysRevB.89.014403}}.
\newline\urlprefix\url{https://link.aps.org/doi/10.1103/PhysRevB.89.014403}

\bibitem{Cebers1975}
A.~O. Tsebers,
  \href{http://mhd.sal.lv/contents/1975/3/MG.11.3.2.R.html}{Simultaneous
  rotational diffusion of the magnetic moment and the solid matrix of a
  single-domain ferromagnetic particle}, Magnetohydrodynamics 11~(3) (1975)
  273--278.
\newline\urlprefix\url{http://mhd.sal.lv/contents/1975/3/MG.11.3.2.R.html}

\bibitem{Mamiya2011}
H.~Mamiya, B.~Jeyadevan,
  \href{http://dx.doi.org/10.1038/srep00157}{Hyperthermic effects of
  dissipative structures of magnetic nanoparticles in large alternating
  magnetic fields}, Scientific Reports 1 (2011) 157 EP, article.
\newblock \href {http://dx.doi.org/10.1038/srep00157}
  {\path{doi:10.1038/srep00157}}.
\newline\urlprefix\url{http://dx.doi.org/10.1038/srep00157}

\bibitem{doi:10.1063/1.4737126}
N.~A. Usov, B.~Y. Liubimov, \href{http://dx.doi.org/10.1063/1.4737126}{Dynamics
  of magnetic nanoparticle in a viscous liquid: Application to magnetic
  nanoparticle hyperthermia}, Journal of Applied Physics 112~(2) (2012) 023901.
\newblock \href {http://arxiv.org/abs/http://dx.doi.org/10.1063/1.4737126}
  {\path{arXiv:http://dx.doi.org/10.1063/1.4737126}}, \href
  {http://dx.doi.org/10.1063/1.4737126} {\path{doi:10.1063/1.4737126}}.
\newline\urlprefix\url{http://dx.doi.org/10.1063/1.4737126}

\bibitem{doi:10.1063/1.4937919}
K.~D. Usadel, C.~Usadel, \href{http://dx.doi.org/10.1063/1.4937919}{Dynamics of
  magnetic single domain particles embedded in a viscous liquid}, Journal of
  Applied Physics 118~(23) (2015) 234303.
\newblock \href {http://arxiv.org/abs/http://dx.doi.org/10.1063/1.4937919}
  {\path{arXiv:http://dx.doi.org/10.1063/1.4937919}}, \href
  {http://dx.doi.org/10.1063/1.4937919} {\path{doi:10.1063/1.4937919}}.
\newline\urlprefix\url{http://dx.doi.org/10.1063/1.4937919}

\bibitem{Usov2015339}
N.~Usov, B.~Y. Liubimov,
  \href{http://www.sciencedirect.com/science/article/pii/S0304885315002565}{Magnetic
  nanoparticle motion in external magnetic field}, Journal of Magnetism and
  Magnetic Materials 385 (2015) 339 -- 346.
\newblock \href
  {http://dx.doi.org/http://dx.doi.org/10.1016/j.jmmm.2015.03.035}
  {\path{doi:http://dx.doi.org/10.1016/j.jmmm.2015.03.035}}.
\newline\urlprefix\url{http://www.sciencedirect.com/science/article/pii/S0304885315002565}

\bibitem{0022-3727-39-22-002}
H.~Xi, K.-Z. Gao, Y.~Shi, S.~Xue,
  \href{http://stacks.iop.org/0022-3727/39/i=22/a=002}{Precessional dynamics of
  single-domain magnetic nanoparticles driven by small ac magnetic fields},
  Journal of Physics D: Applied Physics 39~(22) (2006) 4746.
\newline\urlprefix\url{http://stacks.iop.org/0022-3727/39/i=22/a=002}

\bibitem{PhysRevE.87.062318}
J.~C\ifmmode~\bar{\imath}\else \={\i}\fi{}murs, A.~C\ifmmode~\bar{e}\else
  \={e}\fi{}bers,
  \href{https://link.aps.org/doi/10.1103/PhysRevE.87.062318}{Dynamics of
  anisotropic superparamagnetic particles in a precessing magnetic field},
  Phys. Rev. E 87 (2013) 062318.
\newblock \href {http://dx.doi.org/10.1103/PhysRevE.87.062318}
  {\path{doi:10.1103/PhysRevE.87.062318}}.
\newline\urlprefix\url{https://link.aps.org/doi/10.1103/PhysRevE.87.062318}

\bibitem{PhysRevB.95.134447}
H.~Keshtgar, S.~Streib, A.~Kamra, Y.~M. Blanter, G.~E.~W. Bauer,
  \href{https://link.aps.org/doi/10.1103/PhysRevB.95.134447}{Magnetomechanical
  coupling and ferromagnetic resonance in magnetic nanoparticles}, Phys. Rev. B
  95 (2017) 134447.
\newblock \href {http://dx.doi.org/10.1103/PhysRevB.95.134447}
  {\path{doi:10.1103/PhysRevB.95.134447}}.
\newline\urlprefix\url{https://link.aps.org/doi/10.1103/PhysRevB.95.134447}

\bibitem{C3RA45457F}
M.~I. Dar, S.~A. Shivashankar,
  \href{http://dx.doi.org/10.1039/C3RA45457F}{Single crystalline magnetite{,}
  maghemite{,} and hematite nanoparticles with rich coercivity}, RSC Adv. 4
  (2014) 4105--4113.
\newblock \href {http://dx.doi.org/10.1039/C3RA45457F}
  {\path{doi:10.1039/C3RA45457F}}.
\newline\urlprefix\url{http://dx.doi.org/10.1039/C3RA45457F}

\bibitem{Raikher_1994}
Y.~L. Raikher, M.~I. Shliomis,
  \href{http://dx.doi.org/10.1002/9780470141465.ch8}{The effective field method
  in the orientational kinetics of magnetic fluids and liquid crystals},
  Advances in Chemical Physics 87 (1994) 595--751.
\newblock \href {http://dx.doi.org/http://dx.doi.org/10.1002/9780470141465.ch8}
  {\path{doi:http://dx.doi.org/10.1002/9780470141465.ch8}}.
\newline\urlprefix\url{http://dx.doi.org/10.1002/9780470141465.ch8}

\bibitem{frenkel1955kinetic}
J.~Frenkel, \href{https://books.google.com.ua/books?id=ORdSQwAACAAJ}{Kinetic
  Theory of Liquids}, Dover Publications, Dover, 1955.
\newline\urlprefix\url{https://books.google.com.ua/books?id=ORdSQwAACAAJ}

\end{thebibliography}

\end{document}